\documentclass[preprint,prd,nofootinbib,tightenlines,amsmath,amssymb]{revtex4}

\usepackage{bm}
\usepackage{color}
\usepackage{graphicx}

\begin{document}
\baselineskip=15pt \parskip=5pt

\vspace*{3em}

\title{New LUX and PandaX-II Results Illuminating the Simplest Higgs-Portal Dark Matter Models}

\author{Xiao-Gang He$^{1,2,3}$}
\email{hexg@phys.ntu.edu.tw}
\author{Jusak Tandean$^{2,3}$}
\email{jtandean@yahoo.com}

\affiliation{$^1$INPAC, Department of Physics and Astronomy, Shanghai Jiao Tong University,
800 Dongchuan Rd., Minhang, Shanghai 200240, China \smallskip \\
$^2$Department of Physics and Center for Theoretical Sciences, National Taiwan University,\\
No.\,\,1, Sec.\,\,4, Roosevelt Rd., Taipei 106, Taiwan \smallskip \\
$^3$Physics Division, National Center for Theoretical Sciences,
No.\,\,101, Sec.\,\,2, Kuang Fu Rd., Hsinchu 300, Taiwan \bigskip}


\begin{abstract}
Direct searches for dark matter (DM) by the LUX and PandaX-II Collaborations employing
xenon-based detectors have recently come up with the most stringent limits to date
on the spin-independent elastic scattering of DM off nucleons.
For Higgs-portal scalar DM models, the new results have precluded any possibility of
accommodating low-mass DM as suggested by the DAMA and CDMS II Si experiments utilizing
other target materials, even after invoking isospin-violating DM interactions with nucleons.
In the simplest model, SM+D, which is the standard model plus a real singlet scalar named
darkon acting as the DM candidate, the LUX and PandaX-II limits rule out DM masses
\mbox{roughly from 4 to 450 GeV}, except a small range around the resonance point at half
of the Higgs mass where the interaction cross-section is near the neutrino-background floor.
In the THDM\,II+D, which is the type-II two-Higgs-doublet model combined with a darkon,
the region excluded in the SM+D by the direct searches can be recovered due to suppression
of the DM effective interactions with nucleons at some values of the ratios of Higgs couplings
to the up and down quarks, making the interactions significantly isospin-violating.
However, in either model, if the 125-GeV Higgs boson is the portal between the dark and SM
sectors, DM masses less than 50 GeV or so are already ruled out by the LHC constraint on
the Higgs invisible decay.
In the THDM\,II+D, if the heavier $CP$-even Higgs boson is the portal, theoretical
restrictions from perturbativity, vacuum stability, and unitarity requirements turn out to
be important instead and exclude much of the region below 100 GeV.
For larger DM masses, the THDM\,II+D has plentiful parameter space that corresponds to
interaction cross-sections under the neutrino-background floor and therefore is likely
to be beyond the reach of future direct searches without directional sensitivity.

\end{abstract}

\maketitle

\section{Introduction\label{intro}}

Cosmological studies have led to the inference that ordinary matter makes up only about 5\% of
the energy budget of the Universe, the rest being due to dark matter (26\%) and dark
energy\,\,(69\%), the properties of which are largely still unknown~\cite{pdg}.
Although the evidence for cosmic dark matter (DM) has been established for decades from
numerous observations of its gravitational effects, the identity of its basic constituents has
so far remained elusive.
As the standard model (SM) of particle physics cannot account for the bulk of the DM, it is
of great interest to explore various possible scenarios beyond the SM that can accommodate it.
Amongst the multitudes of DM candidates that have been proposed in the literature, those
classified as weakly interacting massive particles (WIMPs) are perhaps the leading
favorites~\cite{pdg}.
The detection of a WIMP is then essential not only for understanding the nature of the DM
particle, but also for distinguishing models of new physics beyond the SM.

\begin{figure}[t]
\includegraphics[width=83mm]{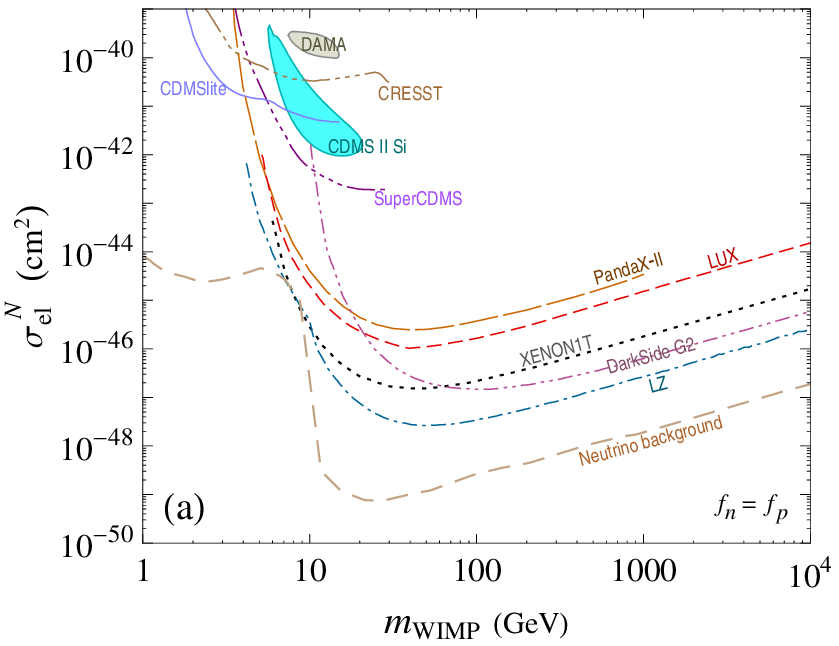} ~
\includegraphics[width=83mm]{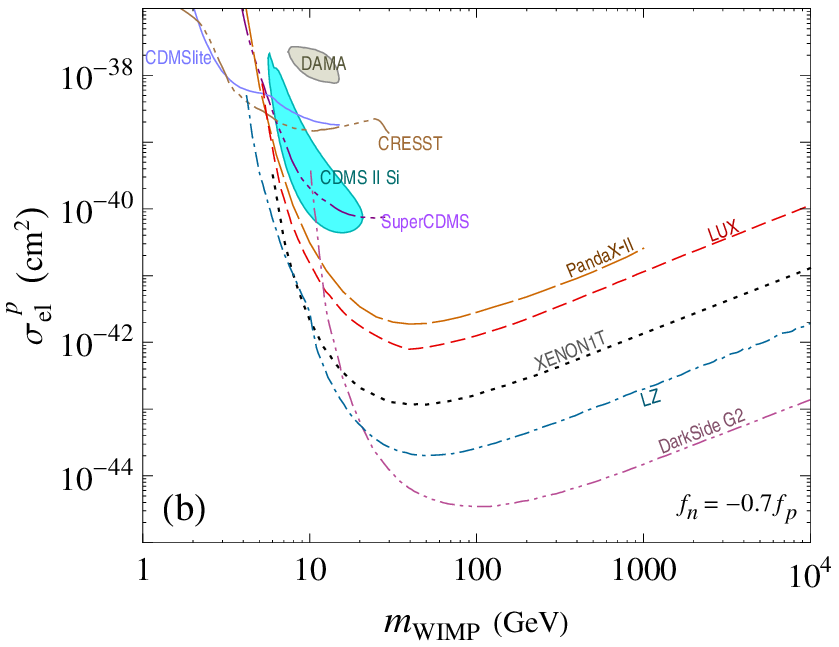}\vspace{-1ex}
\caption{(a) Measured upper-limits on the spin-independent elastic WIMP-nucleon
cross-section at 90\% confidence level (CL) versus WIMP mass from LUX~\cite{lux},
PandaX-II~\cite{pandax}, CDMSlite~\cite{Agnese:2014aze}, SuperCDMS~\cite{Agnese:2015nto},
and CRESST~\cite{Angloher:2015ewa} in the isospin-symmetric limit.
Also shown are a gray patch compatible with the DAMA Na modulation signal at the 3$\sigma$
level~\cite{Savage:2008er}, a cyan area for the possible DM hint from CDMS\,\,II\,\,Si at
90\% CL~\cite{cdmssi}, the sensitivity projections~\cite{Cushman:2013zza} of XENON1T~\cite{xenon1t}
(black dotted curve), DarkSide\,\,G2~\cite{dsg2} (maroon dash-dot-dotted curve), and LZ~\cite{lz}
(turquoise dash-dotted curve), and the WIMP discovery lower-limit due to coherent neutrino
scattering backgrounds~\cite{nubg} (brown dashed curve).
(b) The corresponding WIMP-proton cross-sections computed from (a) with isospin-violating effective
WIMP couplings to the neutron and proton in the ratio \,$f_n^{}/f_p^{}=-0.7$.\label{expt-plots}}
\end{figure}

Many different underground experiments have been and are being performed to detect WIMPs
directly by looking for the signatures of nuclear recoils caused by the collisions between
the DM and nucleons.
The majority of these searches have so far come up empty, leading only to upper bounds on
the cross section $\sigma_{\rm el}^N$ of spin-independent elastic WIMP-nucleon scattering.
Experiments utilizing xenon as the target material have turned out to supply the strictest
bounds to date, especially the newest ones reported separately by the LUX
and PandaX-II Collaborations~\cite{lux,pandax}, under the implicit assumption that the DM
interactions with the proton and neutron respect isospin symmetry.
These null results are in conflict with the tentative indications of WIMP signals observed
earlier at relatively low masses in the DAMA~\cite{dama} and CDMS II Si~\cite{cdmssi} measurements,
which employed nonxenon target materials.\footnote{The excess events previously observed in the
CoGeNT~\cite{cogent} and CRESST-II~\cite{cresst} experiments have recently been demonstrated to
be entirely attributable to underestimated backgrounds instead of DM recoils~\cite{nosignal}.}
A\,\,graphical comparison between the new limits on $\sigma_{\rm el}^N$ from LUX and
PandaX-II and the hypothetical signal regions suggested by DAMA and CDMS II Si is presented in
Fig.\,\,\ref{expt-plots}(a).  It also displays the limits from a few other direct
searches~\cite{Agnese:2014aze,Agnese:2015nto,Angloher:2015ewa}, which were more sensitive
to lighter WIMPs, as well as the expected reaches~\cite{Cushman:2013zza} of the upcoming
XENON1T~\cite{xenon1t}, DarkSide G2~\cite{dsg2}, and LZ~\cite{lz} experiments and an estimate
of the WIMP discovery limit due to coherent neutrino scattering backgrounds~\cite{nubg}.

Mechanisms that may reconcile the incompatible null and positive results of the WIMP DM
direct searches have been suggested over the years.
One of the most appealing proposals stems from the realization that the effective couplings
$f_p^{}$ and $f_n^{}$ of the DM to the proton and neutron, respectively, may be very
dissimilar~\cite{Kurylov:2003ra,Feng:2011vu}.
If such a substantial violation of isospin symmetry occurs, the impact on the detection
sensitivity to WIMP collisions can vary significantly, depending on the target material.
In particular, during the collision process the DM may manifest a\,\,xenophobic behavior
brought about by severe suppression of the collective coupling of the DM to xenon nuclei,
but not necessarily to other nuclei~\cite{Feng:2013fyw}.
This can explain why xenon-based detectors still have not discovered any DM,
but DAMA and CDMS II Si perhaps did.
Numerically, in the xenon case the suppression is the strongest if
\,$f_n^{}/f_p^{}\simeq-0.7$\,~\cite{Feng:2011vu}.
Assuming this ratio and applying it to the pertinent formulas provided
in Ref.\,\cite{Feng:2011vu}, one can translate the data in Fig.\,\,\ref{expt-plots}(a)
into the corresponding numbers for the spin-independent elastic WIMP-proton cross-section,
$\sigma_{\rm el}^p$.
The latter are plotted in Fig.\,\ref{expt-plots}(b), where the curve for DarkSide G2,
which will employ an argon target, is scaled up differently from the curves for the xenon
experiments including LZ.
It is now evident that the conjectured signal regions of DAMA and CDMS II Si are no longer
viable in light of the latest LUX and PandaX-II bounds.\footnote{If the DM-nucleon
scattering is both isospin violating and inelastic, which can happen if a spin-1 particle,
such as a $Z'$ boson, is the portal between the DM and SM particles, it may still be
possible to accommodate the potential hint of low-mass DM from CDMS II Si and evade
the limits from xenon detectors at the same time~\cite{inelastic}.
The inelastic-DM approach has also been proposed to explain the DAMA
anomaly~\cite{Scopel:2015eoh}.\medskip}

Since these new results have reduced further the allowed WIMP parameter space, it is of
interest to investigate what implications they may have for the simplest Higgs-portal
WIMP DM models and how these scenarios may be probed more stringently in the future.
For definiteness, in this paper we focus on the SM+D, which is the SM minimally expanded with
the addition of a\,\,real singlet scalar serving as the DM and dubbed darkon, and on its
two-Higgs-doublet extension of type II, which we call THDM\,II+D.\footnote{There are earlier
studies in the literature on various aspects of the SM plus singlet scalar DM, or a greater
scenario containing the model, in which the scalar was
real~\cite{Silveira:1985rk,sm+reald,Cline:2013gha} or complex~\cite{sm+complexd}.
Two-Higgs-doublet extensions of the SM+D have also been explored
previously~\cite{He:2008qm,He:2011gc,Bird:2006jd,2hdm+d,Drozd:2014yla}.}
Specifically, we look at a number of constraints on these two models not only from the most recent
DM direct searches, but also from LHC measurements on the gauge and Yukawa couplings of the 125-GeV
Higgs boson and on its invisible decay mode, as well as from some theoretical requirements.
We find that in the SM+D the darkon mass region up to {\footnotesize\,$\sim$\,}450 GeV is ruled
out, except a small range near the resonant point at half of the Higgs mass where the DM-nucleon
cross-section is close to the neutrino-background floor.
On the other hand, in the THDM\,II+D the region excluded in the SM+D can be partially recovered
because of suppression of the cross section that happens at some values of the product
\,$\tan\alpha\,\tan\beta$\, or \,$\cot\alpha\,\tan\beta$,\, where $\alpha$ is the mixing angle of
the $CP$-even Higgs bosons and $\tan\beta$ the ratio of vacuum expectation values (VEVs) of
the Higgs doublets.

The structure of the rest of the paper is as follows.
We treat the SM+D in Sec.\,\ref{sec:sm+d} and the THDM\,II+D in Sec.\,\ref{sec:2hdm+d}.
We summarize our results and conclude in Sec.\,\ref{conclusion}.
A couple of appendices contain additional formulas and extra details.

\section{Constraints on SM+D\label{sec:sm+d}}

The darkon, $D$, in the SM+D is a real scalar field and transforms as a singlet under
the gauge group of the SM.
Being the DM candidate, $D$ is stable due to an exactly conserved discrete symmetry, $Z_2$,
under which \,$D\to-D$,\, all the other fields being unaffected.
The renormalizable darkon Lagrangian then has the form~\cite{Silveira:1985rk}
\begin{eqnarray}
{\cal L}_D^{} \,=\, \tfrac{1}{2}\,\partial^\mu D\,\partial_\mu^{}D
- \tfrac{1}{4}\lambda_D^{}D^4 - \tfrac{1}{2}\, m_0^2D^2 - \lambda D^2 H^\dagger H \,,
\end{eqnarray}
where $\lambda_D^{}$,  $m_0^{}$, and $\lambda$  are free parameters and $H$ is the Higgs
doublet containing the physical Higgs field~$h$.
After electroweak symmetry breaking
\begin{eqnarray}
{\cal L}_D^{} \,\supset\, -\frac{\lambda_D^{}}{4}\,D^4
- \frac{\bigl(m_0^2+\lambda v^2\bigr)}{2}\,D^2
- \frac{\lambda}{2}\, D^2\, h^2 - \lambda\, D^2\, h v \,,
\end{eqnarray}
where the second term contains the darkon mass
\,$m_D^{}=\bigl(m^2_0+\lambda v^2\bigr)\raisebox{1pt}{$^{1/2}$}$,\, the last two terms play
an important role in determining the DM relic density, and \,$v\simeq246$\,GeV\, is
the vacuum expectation value (VEV) of $H$.
Clearly, the darkon interactions depend on a small number of free parameters, the relevant
ones here being the darkon-Higgs coupling $\lambda$, which pertains to the relic
density, and the darkon mass $m_D^{}$.

In the SM+D, the relic density results from the annihilation of a darkon pair into SM particles
which is induced mainly by the Higgs-exchange process \,$DD\to h^*\to X_{\textsc{sm}}$,\,
where $X_{\textsc{sm}}$ includes all kinematically allowed final states at the darkon
pair's center-of-mass (c.m.) energy, $\sqrt s$.
If the energy exceeds twice the Higgs mass, \,$\sqrt s>2m_h^{}$,\, the channel \,$DD\to hh$\,
also contributes, which arises from contact and $(s,t,u)$-channel diagrams.
Thus, we can write the cross section $\sigma_{\rm ann}^{}$ of the darkon annihilation into SM
particles as
\begin{eqnarray}
\sigma_{\rm ann}^{} &=& \sigma(DD\to h^*\to X_{\textsc{sm}}) \,+\, \sigma(DD\to hh) \,,
\vphantom{|_{\int}^{}} \nonumber \\ \label{DD2h2sm}
\sigma(DD\to h^*\to X_{\textsc{sm}}) &=&
\frac{4\lambda^2v^2}{\big(m_h^2-s\big)\raisebox{1pt}{$^2$}+\Gamma_h^2m_h^2}~
\frac{\sum_i\Gamma\big(\tilde h\to X_{i,\textsc{sm}}\big)}{\sqrt{s-4m_D^2}} \,, ~~~~ ~~~
X_{\textsc{sm}} \,\neq\, hh \,,
\end{eqnarray}
with $\tilde h$ being a virtual Higgs having the same couplings as the physical $h$ and
an invariant mass equal to $\sqrt s$, and the expression for $\sigma(DD\to hh)$ can be
found in Appendix\,\,\ref{formulas}, which also includes an outline of how $\lambda$
is extracted from the observed abundance of DM.
The resulting values of $\lambda$ can then be tested with constraints from other
experimental information.

In numerical work, we take \,$m_h^{}=125.1$\,GeV,\, based on the current data \cite{lhc:mh},
and correspondingly the SM Higgs width \,$\Gamma_h^{\textsc{sm}}=4.08$\,MeV\, \cite{lhctwiki}.
For \,$m_D^{}<m_h^{}/2$,\, the invisible decay channel \,$h\to DD$\, is open and contributes
to the Higgs' total width \,$\Gamma_h^{}=\Gamma_h^{\textsc{sm}}+\Gamma(h\to DD)$\, in
Eq.\,(\ref{DD2h2sm}), where
\begin{eqnarray} \label{Gh2DD}
\Gamma(h\to DD) \,=\, \frac{\lambda^2v^2}{8\pi m_h^{}}\sqrt{1-\frac{4m_D^2}{m_h^2}} \,.
\end{eqnarray}
The Higgs measurements at the LHC provide information pertinent to this process.
In the latest combined analysis on their Higgs data, the ATLAS and CMS
Collaborations~\cite{atlas+cms} have determined the branching fraction of $h$ decay into
channels beyond the SM to be \,${\cal B}_{\textsc{bsm}}^{\rm exp}=0.00^{+0.16}$,\, which
can be interpreted as setting a cap on the Higgs invisible decay,
\,${\cal B}(h\to\rm invisible)_{\rm exp}<0.16$.\,
Accordingly, we can impose
\begin{eqnarray} \label{lhclimit}
{\cal B}(h\to DD) \,=\, \frac{\Gamma(h\to DD)}{\Gamma_h^{}} \,<\, 0.16 \,,
\end{eqnarray}
which as we will see shortly leads to a major restriction on $\lambda$
for \,$m_D^{}<m_h^{}/2$.\,

Direct searches for DM look for the nuclear recoil effects of DM scattering off a nucleon, $N$.
In the SM+D, this is an elastic reaction, \,$DN\to DN$,\, which is mediated by the Higgs in
the $t$ channel and has a cross section of
\begin{eqnarray} \label{csel}
\sigma_{\rm el}^N \,=\,
\frac{\lambda^2g_{NNh\,}^2m_N^2v^2}{\pi\,\bigl(m_D^{}+m_N^{}\bigr)^2 m_h^4}
\end{eqnarray}
for momentum transfers small relative to $m_h^{}$,
where $g_{NNh}^{}$ is the Higgs-nucleon effective coupling.
Numerically, we adopt \,$g_{NNh}^{}=0.0011$,\, which lies at the low end of our earlier
estimates~\cite{He:2010nt,He:2011gc} and is comparable to other recent
calculations~\cite{Cline:2013gha,Cheng:2012qr}.
The strictest limitations on $\sigma_{\rm el}^N$ to date are supplied by the newest null
findings of LUX~\cite{lux} and PandaX-II~\cite{pandax}.

\begin{figure}[t]
\includegraphics[width=82mm]{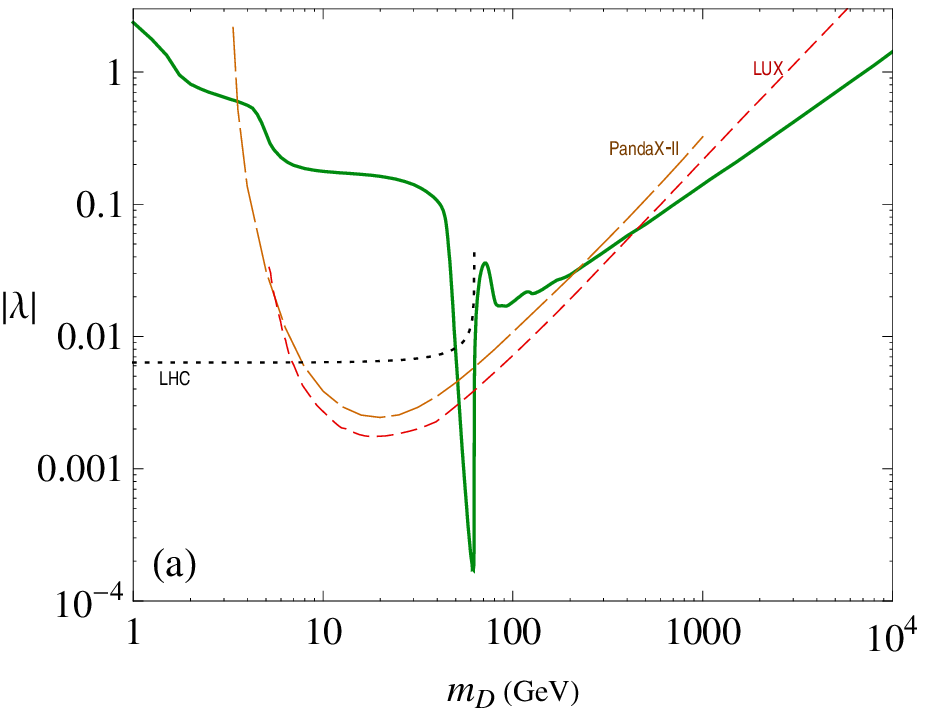} ~~
\includegraphics[width=83mm]{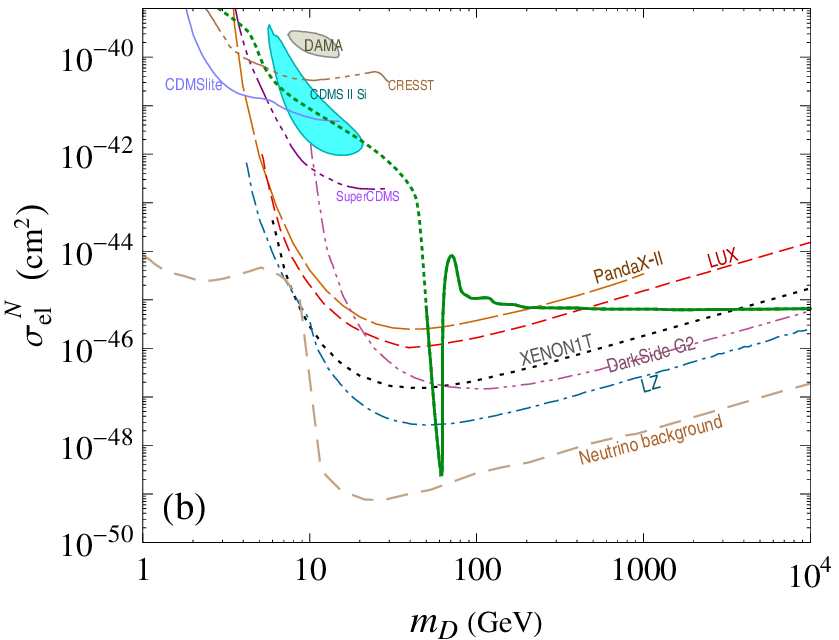}\vspace{-1ex}
\caption{(a) The magnitude of the darkon-Higgs coupling $\lambda$ satisfying the relic abundance
constraint versus the darkon mass $m_D^{}$ in the SM+D (green curve) compared to
the upper limits inferred from LHC data on the Higgs invisible decay (black dotted curve)
and from the latest LUX  (red dashed curve) and PandaX-II (orange dashed curve) searches.
(b) The corresponding darkon-nucleon cross-section~$\sigma_{\rm el}^{N}$ (green curve)
compared to the same current data and future potential limits as in
Fig.\,\,\ref{expt-plots}(a).
The dotted portion of the green curve is excluded by the LHC bound in (a).} \label{sm+d-plots}
\end{figure}

To show how these data confront the SM+D, we display in Fig.\,\ref{sm+d-plots}(a)
the values of $|\lambda|$ derived from the observed relic abundance (green solid curve)
and compare them to the upper bounds on $|\lambda|$ inferred from Eq.\,(\ref{lhclimit})
based on the LHC information on the Higgs invisible decay~\cite{atlas+cms} (black dotted curve)
and from the new results of LUX~\cite{lux} (red dashed curve) and PandaX-II~\cite{pandax}
(orange dashed curve).
The plot in Fig.\,\ref{sm+d-plots}(b) depicts the corresponding prediction for
$\sigma_{\rm el}^N$ (green curve) in comparison to the same DM direct search data and
future potential limits as in Fig.\,\ref{expt-plots}(a).

In the SM+D context, the graphs in Fig.\,\ref{sm+d-plots} reveal that the existing data rule
out darkon masses below about 450 GeV, except for the narrow dip area in the neighborhood
of \,$m_D^{}=m_h^{}/2$,\, more precisely
\,52.1\,\,GeV\,$\mbox{\footnotesize$\lesssim$}\,m_D^{}\,
\mbox{\footnotesize$\lesssim$}\;$62.6\,\,GeV.\,
At \,$m_D^{}=m_h^{}/2$,\, the threshold point for \,$h\to DD$,\, the darkon
annihilation into SM particles undergoes a resonant enhancement, and consequently a\,\,small
size of $\lambda$ can lead to the correct relic density and, at the same time, a low
cross-section of darkon-nucleon collision.
However, as Fig.\,\ref{sm+d-plots} indicates, the bottom of the $\lambda$ dip does not go
to zero due to the Higgs' finite total width $\Gamma_h$ and the annihilation cross-section
at the resonant point being proportional to $1/\Gamma_h^2$.
It is interesting to note that in Fig.\,\ref{sm+d-plots}(b) the bottom of the resonance region
almost touches the expected limit of DM direct detection due to coherent neutrino scattering
backgrounds.
We also notice that the planned XENON1T, DarkSide G2, and LZ experiments~\cite{Cushman:2013zza}
can probe the dip much further, but not all the way down.
Thus, to exclude the dip completely a more sensitive machine will be needed.
For darkon masses above 450 GeV, tests will be available from the ongoing PandaX-II
as well as the forthcoming quests:
particularly, XENON1T, DarkSide G2, and LZ can cover up to
\mbox{\footnotesize\,$\sim$\,}3.5, 10, and a few tens TeV, respectively.

\section{Constraints on THDM\,II+D\label{sec:2hdm+d}}

There are different types of the two-Higgs-doublet model (THDM), depending on how the two
Higgs doublets, $H_1$ and $H_2$, couple to SM fermions~\cite{thdm,Branco:2011iw}.
In the THDM\,\,I, only one of the doublets is responsible for endowing mass to all the fermions.
In the THDM\,\,II, the up-type fermions get mass from only one of the Higgs doublets,
say $H_2$, and the down-type fermions from the other doublet.
In the THDM\,\,III, both $H_1$ and $H_2$ give masses to all the fermions.

Since only one Higgs doublet generates all of the fermion masses in the THDM I, the couplings of
each of the $CP$-even Higgs bosons to fermions are the same as in the SM,
up to an overall scaling factor.
Therefore, the couplings of the 125-GeV Higgs, $h$, in the THDM I slightly enlarged with
the addition of a darkon are similar to those in the SM+D treated in the previous section,
and consequently for \,$m_D^{}<m_h^{}/2$\, the modifications cannot readily ease
the restraints from the DM direct searches and LHC quest for the Higgs invisible decay.
Combining a\,\,darkon with the THDM III instead could provide the desired ingredients to help
overcome these obstacles~\cite{He:2011gc}, but the model possesses too many parameters to be
predictable, some of which give rise to undesirable flavor-changing neutral-Higgs transitions
at tree level.
For these reasons, in the remainder of the section we concentrate on the THDM\,\,II plus
the darkon (THDM\,II+D).

In the THDM\,II+D, the fermion sector is no different from that in the THDM\,II,
with the Yukawa interactions being described by~\cite{thdm,Branco:2011iw}
\begin{eqnarray} \label{LY2hdm}
{\cal L}_{\rm Y}^{} \,=\,
-\overline{Q}_{j,L}^{} \big(\lambda_2^u\big)_{jl} \tilde H_{2\,}^{}{\cal U}_{l,R}^{}
- \overline{Q}_{j,L}^{} \big(\lambda_1^d\big)_{jl} H_1^{} {\cal D}_{l,R}^{}
- \overline{L}_{j,L}^{} \big(\lambda_1^\ell\big)_{jl} H_1^{} E_{l,R}^{}
\;+\; {\rm H.c.} \,,
\end{eqnarray}
where summation over \,$j,l=1,2,3$\, is implicit, $Q_{j,L}$ $(L_{j,L})$ represents left-handed
quark (lepton) doublets, \,${\cal U}_{l,R}^{}$ and ${\cal D}_{l,R}$ $(E_{l,R})$ denote
right-handed quark (charged lepton) fields, \,$\tilde H_{1,2}^{}=i\tau_2^{}H_{1,2}^*$\,
with $\tau_2^{}$ being the second Pauli matrix, and $\lambda^{u,d,\ell}$ are 3$\times$3
matrices for the Yukawa couplings.
This Lagrangian respects the discrete symmetry, $Z_2$, under which \,$H_2\to-H_2$\, and
\,${\cal U}_R\to-{\cal U}_R$,\, while all the other fields are not affected.
Thus, $Z_2$ prohibits the combinations \,$\overline{Q}{}_L\tilde{H}_1{\cal U}_R$,
$\overline{Q}{}_LH_2{\cal D}_R$, $\overline{L}{}_LH_2E_R$, and their Hermitian conjugates from
occurring in ${\cal L}_{\rm Y}$.

The longevity of the darkon as the DM in the THDM\,II+D is maintained by another discrete
symmetry, $Z_2'$, under which \,$D\to-D$,\, whereas all the other fields are $Z_2'$ even.
Consequently, being a real field and transforming as a singlet under the SM gauge group,
$D$ has no renormalizable interactions with SM fermions or gauge bosons, like in the SM+D.

The renormalizable Lagrangian of the model, \,${\cal L}\supset-{\cal V}_D-{\cal V}_H$,\,
contains the scalar potential terms~\cite{Bird:2006jd}
\begin{eqnarray}
{\cal V}_D^{} &=& \frac{m_0^2}{2}\, D^2 + \frac{\lambda_D^{}}{4}\, D^4 \,+\,
\big(\lambda_{1D\,}^{}H_1^\dagger H_1^{}+\lambda_{2D\,}^{}H_2^\dagger H_2^{}\big)D^2 \,,
\nonumber \vphantom{|_{\int_\int^{}}^{}} \\ \label{pot}
{\cal V}_H^{} &=& m_{11\,}^2 H_1^\dagger H_1^{} + m_{22\,}^2 H_2^\dagger H_2^{}
- \big(m_{12}^2\,H_1^\dagger H_2^{}\,+\,{\rm H.c.}\big) \,+\,
\frac{\lambda_1}{2} \bigl(H_1^\dagger H_1^{}\bigr)\raisebox{1pt}{$^2$} +
\frac{\lambda_2}{2} \bigl(H_2^\dagger H_2^{}\bigr)\raisebox{1pt}{$^2$} ~~~~~
\nonumber \\ && \! +\;
\lambda_{3\,}^{}H_1^\dagger H_{1\,}^{}H_2^\dagger H_2^{}
+ \lambda_{4\,}^{}H_1^\dagger H_{2\,}^{}H_2^\dagger H_1^{}
+ \frac{\lambda_5^{}}{2}
\Big[ \big(H_1^\dagger H_2^{}\big)\raisebox{1pt}{$^2$} \,+\,{\rm H.c.}\Big] \,,
\end{eqnarray}
where ${\cal V}_H$ is the usual THDM\,\,II potential~\cite{thdm,Branco:2011iw}.
Because of $Z_2$, the combinations $H_1^\dagger H_2^{}D^2$,
$H_1^\dagger H_{1\,}^{}H_1^\dagger H_2^{}$, $H_1^\dagger H_{2\,}^{}H_2^\dagger H_2^{}$,
and their Hermitian conjugates are forbidden from appearing in Eq.\,(\ref{pot}).
However, in ${\cal V}_H$ we have included the $m_{12}^2$ terms which softly break $Z_2$
and are important in relaxing the upper bounds on the Higgs masses~\cite{Branco:2011iw}.
In contrast, $Z_2'$, which guarantees the darkon stability, is exactly conserved.
The Hermiticity of ${\cal V}_{D,H}$ implies that the parameters $m_{0,11,22}^2$ and
$\lambda_{D,1D,2D,1,2,3,4}$ are real.
We assume ${\cal V}_{D,H}$ to be $CP$ invariant, and so $m_{12}^2$ and $\lambda_5^{}$
are also real parameters.

The $\lambda_{1D,2D}$ terms in Eq.\,(\ref{pot}) play a crucial role in the determination
of the relic density, which follows from darkon annihilation into the other particles
via interactions with the Higgs bosons.
To address this in more detail, we first decompose the Higgs doublets as
\begin{eqnarray}
H_r^{} \,=\, \frac{1}{\sqrt2} \left(\begin{array}{c} \sqrt2\,h_r^+ \vspace{1ex} \\
v_r^{}+ h_r^0 + i I_r^0 \end{array}\right ) \,, ~~~~~
r \,=\, 1,2 \,,
\end{eqnarray}
where $v_{1,2}^{}$ are the VEVs of $H_{1,2}$, respectively, and connected to the electroweak
scale \,$v\simeq246$\,GeV\, by \,$v_1^{}=v\cos\beta$\, and \,$v_2^{}=v\sin\beta$.\,
The $H_r$ components $h_r^+$, $h_r^0$, and $I_r^0$ are related to the physical Higgs
bosons $h$, $H$, $A$, and $H^+$ and the would-be Goldstone bosons $w^+$ and $z$ by
\begin{eqnarray}
\left(\begin{array}{c} h^+_1 \vspace{1ex} \\ h^+_2 \end{array}\right) &\!=\!&
\left(\begin{array}{lr} c_\beta^{}~ & -s_\beta^{} \vspace{3pt} \\
s_\beta^{} & c_\beta^{} \end{array}\right)
\left(\begin{array}{c} w^+ \vspace{1ex} \\ H^+ \end{array}\right) , ~~~~ ~~~
\left(\begin{array}{c} I_1^0 \vspace{1ex} \\ I_2^0 \end{array}\right) = \left(\begin{array}{lr}
c_\beta^{}~ & -s_\beta^{} \vspace{3pt} \\ s_\beta^{} & c_\beta^{} \end{array}\right)
\left(\begin{array}{c} z \vspace{3pt} \\ A \end{array}\right) ,
\nonumber \\
\left(\begin{array}{c} h_1^0 \vspace{1ex} \\ h_2^0 \end{array}\right) &\!=\!&
\left(\begin{array}{lr} c_\alpha^{}~ & -s_\alpha^{} \vspace{3pt} \\
s_\alpha^{} & c_\alpha^{} \end{array}\right)
\left(\begin{array}{c} H \vspace{3pt} \\ h \end{array}\right) , \hspace{4.7em}
c_{\cal X}^{} \,=\, \cos{\cal X} \,, ~~~~~ s_{\cal X}^{} \,=\, \sin{\cal X} \,,
\end{eqnarray}
where $\cal X$ is any angle or combination of angles.
The $w^\pm$ and $z$ will be eaten by the $W^\pm$ and $Z$ bosons, respectively.

After electroweak symmetry breaking, we can then express the relevant terms in
\,${\cal V}={\cal V}_D+{\cal V}_H$\, involving the physical bosons as
\begin{eqnarray} \label{LHD}
{\cal V} &\supset&
\tfrac{1}{2}\,m_D^2 D^2 \,+\, \big(\lambda_h^{} h + \lambda_H^{} H\big)D^2 v
\nonumber \\ && \!\! +\;
\tfrac{1}{2} \big( \lambda_{hh}^{}h^2 + 2\lambda_{hH\,}^{}h H
+ \lambda_{HH}^{}H^2 + \lambda_{AA}^{}A^2 + 2\lambda_{H^+H^-}^{}H^+H^- \big) D^2
\nonumber \\ && \!\! +\;
\Big( \tfrac{1}{6} \lambda_{hhh}^{}h^2 + \tfrac{1}{2} \lambda_{hhH}^{}h H
+ \tfrac{1}{2} \lambda_{hHH}^{}H^2 + \tfrac{1}{2} \lambda_{hAA}^{}A^2
+ \lambda_{hH^+H^-}^{}H^+H^- \Big) h v
\nonumber \\ && \!\! +\;
\Big( \tfrac{1}{6} \lambda_{HHH}^{}H^2 + \tfrac{1}{2} \lambda_{HAA}^{}A^2
+ \lambda_{HH^+H^-}^{}H^+H^- \Big) H v \,,
\end{eqnarray}
where \,$m_D^2=m_0^2+\bigl(\lambda_{1D}^{}\,c_\beta^2+\lambda_{2D}^{}\,s_\beta^2\bigr)v^2$,\,
\begin{eqnarray} \label{lambdah}
\lambda_h^{} &=&
\lambda_{2D}^{}\,c_\alpha^{}s_\beta^{} - \lambda_{1D}^{}\,s_\alpha^{}c_\beta^{} \,, \hspace{5.7em}
\lambda_H^{} \,=\,
\lambda_{1D}^{}\,c_\alpha^{}c_\beta^{} + \lambda_{2D}^{}\,s_\alpha^{}s_\beta^{} \,,
\nonumber \\
\lambda_{hh}^{} &=& \lambda_{1D}^{}\,s_\alpha^2 + \lambda_{2D}^{}\,c_\alpha^2 \,, \hspace{6.9em}
\lambda_{HH}^{} \,=\, \lambda_{1D}^{}\,c_\alpha^2 + \lambda_{2D}^{}\,s_\alpha^2 \,,
\nonumber \\
\lambda_{hH}^{} &=& \big(\lambda_{2D}^{}-\lambda_{1D}^{}\big)c_\alpha^{}s_\alpha^{} \,,
\hspace{6.5em} \lambda_{AA}^{} \,=\, \lambda_{H^+H^-}^{} \,=\,
\lambda_{1D}^{}\,s_\beta^2 \,+\, \lambda_{2D}^{}\,c_\beta^2 \,,
\end{eqnarray}
and the cubic couplings $\lambda_{\texttt{XYZ}}$ are listed in Appendix\,\,\ref{formulas}.
There is no $AD^2$ term under the assumed $CP$ invariance.
Since $m_0^{}$ and $\lambda_{1,2}$ are free parameters, so are $m_D^{}$ and~$\lambda_{h,H}$.
The quartic couplings of the darkon to the Higgs bosons can then be related to $\lambda_{h,H}$ by
\begin{eqnarray} \label{lambdahh}
\lambda_{hh}^{} &=&\!
\Bigg(\frac{c_\alpha^3}{s_\beta^{}}-\frac{s_\alpha^3}{c_\beta^{}}\Bigg)\lambda_h^{}
+ \frac{s_{2\alpha}^{}c_{\beta-\alpha}^{}}{s_{2\beta}^{}}\,\lambda_H^{} \,, \hspace{9ex}
\lambda_{HH}^{} \,=
\Bigg( \frac{c_\alpha^3}{c_\beta^{}}+\frac{s_\alpha^3}{s_\beta^{}} \Bigg) \lambda_H^{}
- \frac{s_{2\alpha}^{}s_{\beta-\alpha}^{}}{s_{2\beta}^{}}\,\lambda_h^{} \,,
\nonumber \\
\lambda_{hH}^{} &=&
\frac{s_{2\alpha}^{}}{s_{2\beta}^{}}\big(\lambda_h^{}c_{\beta-\alpha}^{}
- \lambda_H^{}s_{\beta-\alpha}^{} \big) \,, \hspace{4ex}
\lambda_{AA}^{} \,=\, \lambda_{H^+H^-}^{} \,=\,
\frac{c_\alpha^{}c_\beta^3-s_\alpha^{}s_\beta^3}{c_\beta^{}s_\beta^{}}\lambda_h^{} +
\frac{c_\alpha^{}s_\beta^3+s_\alpha^{}c_\beta^3}{c_\beta^{}s_\beta^{}}\lambda_H^{} \,. ~~~~
\end{eqnarray}

Since $h$ and $H$ couple directly to the weak bosons, we need to include the annihilation
channels \,$DD\to W^+W^-,ZZ$\, if kinematically permitted.
The pertinent interactions are given by
\begin{eqnarray} \label{vvh}
{\cal L} \,\supset\, \bigl(2m_W^2W^{+\mu}W_\mu^-+m_Z^2Z^\mu Z_\mu^{}\bigr)
\bigg(k_V^h\,\frac{h}{v}+k_V^H\,\frac{H}{v}\bigg) \,, ~~~~ ~~~
k_V^h \,=\, s_{\beta-\alpha}^{} \,, ~~~~~ k_V^H \,=\, c_{\beta-\alpha}^{} \,. ~~
\end{eqnarray}

The scattering of the darkon off a nucleon \,${\cal N}=p$ or $n$\, is generally mediated
at the quark level by $h$ and $H$ and hence depends not only on the darkon-Higgs couplings
$\lambda_{h,H}$, but also on the effective Higgs-nucleon coupling\,\,$g_{\cal NNH}^{}$
defined by
\begin{eqnarray}
{\cal L}_{\cal NNH}^{} \,=\, -g_{\cal NNH}^{}\, \overline{\cal N}{\cal N}{\cal H} \,, ~~~ ~~~~
{\cal H} \,=\, h,H \,.
\end{eqnarray}
This originates from the quark-Higgs terms in Eq.\,(\ref{LY2hdm}) given by
\begin{eqnarray}
{\cal L}_{\rm Y}^{} \,\supset\, -\raisebox{-7pt}{\Large$\stackrel{\sum}{\mbox{\scriptsize$q$}}$}\,
k_q^{\cal H}m_q^{}\,\overline{q}q\,\frac{\cal H}{v} \,, ~~~~ ~~~
k_{c,t}^{\cal H} \,=\, k_u^{\cal H} \,, ~~~~~ k_{s,b}^{\cal H} \,=\, k_d^{\cal H} \,, ~~~~
\end{eqnarray}
where the sum is over all quarks, \,$q=u,d,s,c,b,t$,\, and
\begin{eqnarray}  \label{kukdII}
k_u^h \,=\,  \frac{c_\alpha^{}}{s_\beta^{}} \,, ~~~~ ~~~
k_d^h \,=\, -\frac{s_\alpha^{}}{c_\beta^{}} \,, ~~~~ ~~~
k_u^H \,=\,  \frac{s_\alpha^{}}{s_\beta^{}} \,, ~~~~ ~~~
k_d^H \,=\,  \frac{c_\alpha^{}}{c_\beta^{}} \,.
\end{eqnarray}
It follows that~\cite{Shifman:1978zn}
\begin{eqnarray} \label{gNNH}
g_{\cal NNH}^{} \,=\, \frac{m_{\cal N}^{}}{v}
\Big[ \Big(f_u^{\cal N}+f_c^{\cal N}+f_t^{\cal N}\Big) k_u^{\cal H}
+ \Big(f_d^{\cal N}+f_s^{\cal N}+f_b^{\cal N}\Big) k_d^{\cal H} \Big] \,,
\end{eqnarray}
where $f_q^{\cal N}$ is defined by the matrix element
\,$\langle{\cal N}|m_q^{}\overline{q}q|{\cal N}\rangle=
f_q^{\cal N}m_{\cal N\,}^{}\overline{u}_{\cal N}^{}u_{\cal N}^{}$\,
with $u_{\cal N}$ being the Dirac spinor for $\cal N$ and $m_{\cal N}^{}$ its mass.
Employing the values $f_q^{\cal N}$ for the different quarks listed in
Appendix\,\,\ref{formulas}, we find
\begin{eqnarray} \label{gNNH2hdm}
g_{pp\cal H}^{} \,=\, \big(0.5631\,k_u^{\cal H}+0.5599\,k_d^{\cal H}\big)\times10^{-3} \,, ~~~~~
g_{nn\cal H}^{} \,=\, \big(0.5481\,k_u^{\cal H}+0.5857\,k_d^{\cal H}\big)\times10^{-3} \,. ~~~
\end{eqnarray}
Setting \,$k_{u,d}^h=1$\, in these formulas, we reproduce the SM values
\,$g_{pph,nnh}^{\textsc{sm}}\simeq0.0011$\, quoted in the last section.
However, if $k_{u,d}^{\cal H}$ are not close to unity, $g_{pp\cal H}^{}$ and
$g_{nn\cal H}^{}$ can be very dissimilar, breaking isospin symmetry substantially.
Particularly, they have different zeros, \,$k_d^{\cal H}\simeq-1.01\,k_u^{\cal H}$\,
and \,$k_d^{\cal H}\simeq-0.936\,k_u^{\cal H}$,\, respectively.

\begin{figure}[t]
\includegraphics[width=8cm]{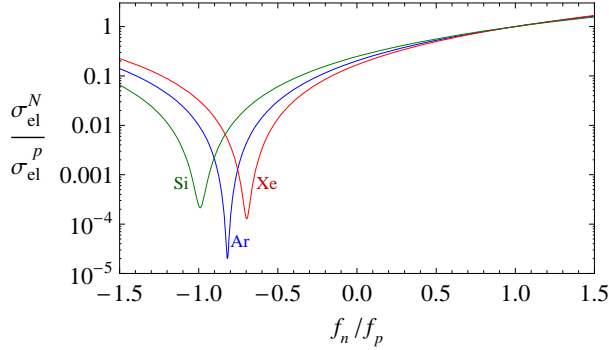}\vspace{-1ex}
\caption{Dependence of $\sigma_{\rm el}^N/\sigma_{\rm el}^p$ on $f_n^{}/f_p^{}$ according to
Eq.\,(\ref{ivdm}) for silicon, argon, and xenon targets.\label{sigmaratio}}
\end{figure}

This suggests that to evaluate DM collisions with nucleons in the THDM\,II+D it is more
appropriate to work with either the darkon-proton or darkon-neutron cross-section
($\sigma_{\rm el}^p$ or $\sigma_{\rm el}^n$, respectively) rather than the darkon-nucleon
one under the assumption of isospin conservation.
The calculated $\sigma_{\rm el}^{p,n}$ can then be compared to their empirical
counterparts which are converted from the measured $\sigma_{\rm el}^N$ using
the relations~\cite{Feng:2011vu,Feng:2013fyw}
\begin{eqnarray} \label{ivdm}
\sigma_{\rm el}^N\,\raisebox{-8pt}{\Large$\stackrel{\sum}{\mbox{\scriptsize$i$}}$}\,
\eta_i^{}\,\mu_{A_i}^2A_i^2 \,=\,
\sigma_{\rm el}^p\,\raisebox{-8pt}{\Large$\stackrel{\sum}{\mbox{\scriptsize$i$}}$}\,\eta_i^{}\,
\mu_{A_i}^2\big[{\cal Z}+\big(A_i^{}-{\cal Z}\bigr)f_n^{}/f_p^{}\big]\raisebox{1pt}{$^2$} \,,
~~~~ ~~~ \sigma_{\rm el}^n \,=\, \sigma_{\rm el}^p\,f_n^2/f_p^2 \,,
\end{eqnarray}
where the sums are over the isotopes of the element in the target material with which the DM
interacts dominantly, $\eta_i^{}$  $(A_i)$ represent the fractional abundances
(the nucleon numbers) of the isotopes,\footnote{A recent list of isotopic abundances
can be found in~\cite{isotopes}.} \,$\mu_{A_i}=m_{A_i}m_D/\bigl(m_{A_i}+m_D^{}\bigr)$,\,
with $m_{A_i}$ being the $i$th isotope's mass, $\cal Z$ denotes the proton number of
the element, and $f_n^{}/f_p^{}$ is fixed under certain assumptions.
For illustration, from Eq.\,(\ref{ivdm}) we graph $\sigma_{\rm el}^N/\sigma_{\rm el}^p$ as
a function $f_n^{}/f_p^{}$ for a few target materials (silicon, argon, and xenon) in
Fig.\,\ref{sigmaratio}, where the curves are not sensitive to the darkon masses in our
range of interest.
Thus, if there is no isospin violation, \,$f_n^{}=f_p^{}$\, leading to
\,$\sigma_{\rm el}^p=\sigma_{\rm el}^N$.\,
On the other hand, for DM with maximal xenophobia, \,$f_n^{}/f_p^{}=-0.70$,\, and with this
number we arrived at Fig.\,\ref{expt-plots}(b) from Fig.\,\ref{expt-plots}(a).
More generally, $\sigma_{\rm el}^p$ can be bigger or smaller than $\sigma_{\rm el}^N$\,
if \,$f_n^{}\neq f_p^{}$,\, but completely destructive interference on the right-hand side of
the first relation in Eq.\,(\ref{ivdm}) yielding \,$\sigma_{\rm el}^N/\sigma_{\rm el}^p=0$\,
is not achievable if the element has more than one naturally abundant isotope.

If both the $h$ and $H$ couplings to the darkon are nonzero, the cross section of
the darkon-$\cal N$ scattering \,$D{\cal N}\to D{\cal N}$\, is
\begin{eqnarray} \label{DN2DN}
\sigma_{\rm el}^{\cal N} \,=\, \frac{m_{\cal N}^2\,v^2}{\pi\,\bigl(m_D^{}+m_{\cal N}^{}\bigr)^2}
\Biggl( \frac{\lambda_h^{}\,g_{{\cal NN}h}^{}}{m_h^2} +
\frac{\lambda_H^{}\,g_{{\cal NN}H}^{}}{m_H^2} \Biggr)^{\!\!2}
\end{eqnarray}
for momentum transfers small relative to $m_{h,H}^{}$ and \,${\cal N}=p$ or $n$.\,
Given that $g_{\cal NNH}^{}$ depends on $k_{u,d}^{\cal H}$ according to Eq.\,(\ref{gNNH}),
it may be possible to make $g_{\cal NNH}^{}$ sufficiently small with a suitable choice of
$k_d^{\cal H}/k_u^{\cal H}$ to allow $\sigma_{\rm el}^{\cal N}$ to avoid its experimental
limit~\cite{He:2008qm}, at least for some of the $m_D^{}$ values.
Moreover, the $\lambda_{h,H}$ terms in Eq.\,(\ref{DN2DN}) may (partially) cancel each other
to reduce $\sigma_{\rm el}^{\cal N}$ as well.
These are attractive features of the THDM\,II+D that the SM+D does not possess.

Since there are numerous different possibilities in which $h$ and $H$ may contribute to darkon
interactions with SM particles in the THDM+D, hereafter for definiteness and simplicity we
focus on a couple of scenarios in which $h$ is the 125-GeV Higgs boson and the other Higgs
bosons are heavier, \,$m_h^{}<m_{H,A,H^\pm}$.\,
In addition, we assume specifically that either $H$ or $h$ has a\,\,vanishing coupling to
the darkon, \,$\lambda_H=0$\, or \,$\lambda_h=0$,\, respectively.
As a consequence, either $h$ or $H$ alone serves as the portal between the DM and SM
particles, and so we now have \,$f_n^{}/f_p^{}=g_{nn\cal H}^{}/g_{pp\cal H}^{}$,\,
upon neglecting the $n$-$p$ mass difference.

If we take \,$g_{nn\cal H}^{}/g_{pp\cal H}^{}=-0.70$,\, which corresponds to the xenophobic limit,
using Eq.\,(\ref{gNNH2hdm}) we get \,$r_k^{\cal H}\equiv k_d^{\cal H}/k_u^{\cal H}=-0.96$,\,
where \,$r_k^h=-\tan\alpha\,\tan\beta$\, and \,$r_k^H=\cot\alpha\,\tan\beta$\,
from Eq.\,(\ref{kukdII}).
Nevertheless, as we see later on, despite the strongest constraints to date from
xenon-based detectors, higher $r_k^{\cal H}$ values are still compatible with
the data and hence the darkon can still avoid extreme xenophobia.
The choices for $\alpha$ and $\beta$, however, need to comply with further restraints
on $k_{d,u,V}^h$, as specified below.

\begin{figure}[h]
\includegraphics[width=54mm]{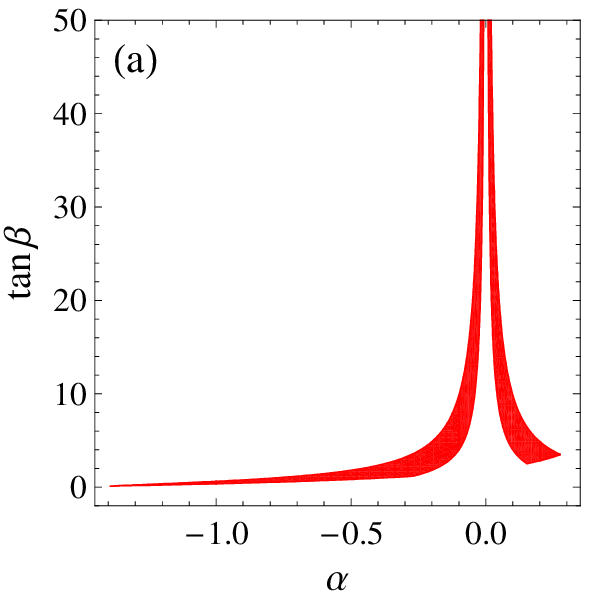} ~ ~
\includegraphics[width=53mm]{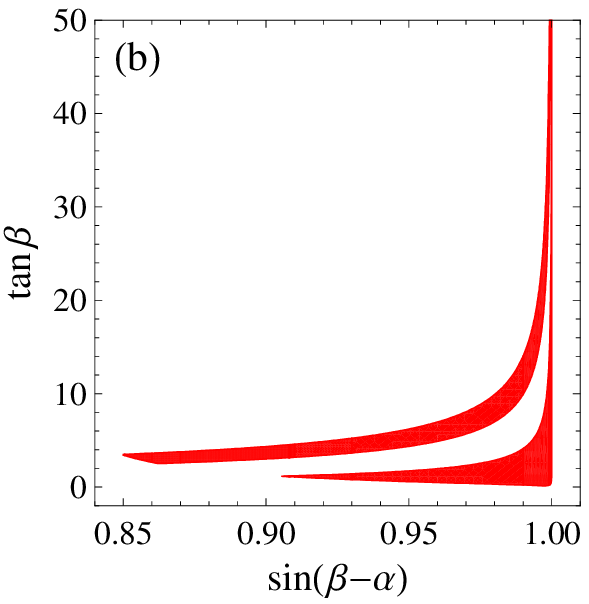}\vspace{-1ex}
\caption{Regions of $\tan\beta$ versus (a) $\alpha$ and (b) $\sin(\beta-\alpha)$
which obey the restrictions in Eq.\,(\ref{khconstr}).\label{tanb-a}}
\end{figure}

Given that LHC measurements have been probing the Higgs couplings to SM fermions
and electroweak bosons, we need to take into account the resulting restrictions on
potential new physics in the couplings.
A modification to the \,$h\to X\bar X$\, interaction with respect to its SM
expectation can be parameterized by $\kappa_X$ defined by
\,$\kappa_X^2=\Gamma_{h\to X\bar X}^{}/\Gamma_{h\to X\bar X}^{\textsc{sm}}$.\,
Assuming that \,$|\kappa_{W,Z}|\le1$\, and the Higgs total width can get contributions
from decay modes beyond the SM, the ATLAS and CMS Collaborations have performed
simultaneous fits to their Higgs data to extract~\cite{atlas+cms}
\begin{eqnarray} \label{kappaex}
\kappa_W^{} &=& 0.90\pm0.09 \,, ~~~ ~~~~ \kappa_t^{} \,=\, 1.43_{-0.22}^{+0.23} \,, ~~~ ~~~~
|\kappa_b^{}| \,=\, 0.57\pm0.16 \,, ~~~ ~~~~
|\kappa_\gamma| \,=\, 0.90_{-0.09}^{+0.10} \,,
\nonumber \\
\kappa_Z^{} &=& 1.00_{-0.08}^{} \,, ~~~~ ~~~~\, |\kappa_g^{}| \,=\, 0.81_{-0.10}^{+0.13} \,,
~~~ ~~~\, |\kappa_\tau^{}| \,=\, 0.87_{-0.11}^{+0.12} \,,
\end{eqnarray}
where~\cite{atlas+cms}
\,$\kappa_\gamma^2=0.07\,\kappa_t^2+1.59\,\kappa_W^2-0.66\,\kappa_t^{}\kappa_W^{}$.\,
In the THDM\,\,II context, we expect these numbers to respect within one sigma the relations
\,$k_V^h=\kappa_W^{}=\kappa_Z^{}$,\, $k_u^h=\kappa_t^{}\simeq\kappa_g^{}$,\, and
\,$k_d^h=\kappa_b^{}=\kappa_\tau^{}$,\, although the $\kappa_{t,g}^{}$ $(\kappa_{b,\tau})$
numbers above overlap only at the two-sigma level.
Accordingly, pending improvement in the precision of these parameters from future data,
based on Eq.\,(\ref{kappaex}) we may impose
\begin{eqnarray} \label{khconstr}
0.81 \,\le\, k_V^h \,\le\, 1 \,, ~~~~~ 0.71 \,\le\, k_u^h \,\le\, 1.66 \,, ~~~~~
0.41 \,\le\, \big|k_d^h\big| \,\le\, 0.99 \,, ~~~~~
0.81 \,\le\, \big|k_\gamma^h\big| \,\le\, 1 \,, ~~~
\end{eqnarray}
where $k_\gamma^h$ incorporates the loop contribution of $H^\pm$ to
\,$h\to\gamma\gamma$,\, and so \,$k_\gamma^h\to\kappa_\gamma^{}$\, if the impact
of $H^\pm$ is vanishing.
Explicitly
\begin{eqnarray} \label{kgamma}
k_\gamma^h \,=\, 0.264\,k_u^h \,-\, 1.259\,k_V^h \,+\,
0.151\;\frac{\lambda_{hH^+H^-}^{}v^2}{2m_{H^\pm}^2}\,
A_0^{\gamma\gamma}\big(4m_{H^\pm}^2/m_h^2\big) \,,
\end{eqnarray}
where $A_0^{\gamma\gamma}$ is a loop function whose expression can be found in
the literature ({\it e.g.}, \cite{Chen:2013vi}).
The effect of the $\lambda_{hH^+H^-}$ term in $k_\gamma^h$ turns out to be somewhat
minor in our examples.
To visualize the impact of the limitations in Eq.\,(\ref{khconstr}), we plot in
Fig.\,\ref{tanb-a} the (red) regions representing the $\alpha$ and $\beta$
parameter space satisfying them.

Before proceeding to our specific scenarios of choice, we remark that in the alignment
limit, \,$\beta=\alpha+\pi/2$,\, we recover the SM+D darkon parameters,
\begin{eqnarray}
m_D^2 \,=\, m_0^2 + \lambda_h^{}v^2 \,, ~~~~ ~~~ \lambda_{hh}^{} \,=\, \lambda_h^{}
\end{eqnarray}
with \,$\lambda_h=\lambda$.\,
Furthermore, in this limit the $h$ couplings become SM-like,
\begin{eqnarray}
\lambda_{hhh}^{} \,=\, \frac{3m_h^2}{v^2} \,, ~~~~ ~~~
k_V^h \,=\, 1 \,, ~~~~ ~~~ k_q^h \,=\, 1 \,.
\end{eqnarray}

\subsection{\boldmath$\lambda_H=0$\label{0lH}}

In this case, the cross section of the darkon annihilation into THDM particles is
\begin{eqnarray}
\sigma_{\rm ann}^{} \,=\, \sigma(DD\to h^*\to X_{\textsc{sm}}) \,+\,
\raisebox{-7pt}{\Large$\stackrel{\sum}{\mbox{\scriptsize${\texttt s}_1{\texttt s}_2$}}$}\,
\sigma(DD\to{\texttt s}_1{\texttt s}_2) \,,
\end{eqnarray}
where the first term on the right-hand side is equal to its SM+D counterpart in
Eq.\,(\ref{DD2h2sm}), except $\lambda$ is replaced by $\lambda_h$ and the $h$ couplings to
fermions and gauge bosons are multiplied by the relevant $k_{u,d,V}^h$ factors mentioned
earlier, and the sum is over \,${\texttt s}_1{\texttt s}_2=hh,hH,HH,AA,H^+H^-$\,
with only kinematically allowed channels contributing.
The formulas for $\sigma(DD\to{\texttt s}_1{\texttt s}_2)$ have been relegated
to Appendix\,\,\ref{formulas}.
Hence, though not the portal between the DM and SM particles in this scenario, $H$ can still
contribute to the darkon relic abundance via \,$DD\to{\texttt s}_1{\texttt s}_2$,\,
along with $A$ and $H^\pm$.

Once $\lambda_h$ has been extracted from the relic density data and $g_{{\cal NN}h}^{}$
calculated with the $\alpha$ and $\beta$ choices consistent with Eq.\,(\ref{kukdII}),
we can predict the darkon-$\cal N$ cross-section.
From now on, we work exclusively with the darkon-proton one,
\begin{eqnarray} \label{Dp2Dp}
\sigma_{\rm el}^p \,=\, \frac{\lambda_h^2\,g_{pph}^2\,m_p^2\,v^2}
{\pi\,\bigl(m_D^{}+m_p^{}\bigr)\raisebox{1pt}{$^2$}\,m_h^4} \,.
\end{eqnarray}
This is to be compared to its empirical counterparts derived from the $\sigma_{\rm el}^N$
data using Eq.\,(\ref{ivdm}) with \,$f_n^{}/f_p^{}=g_{nnh}^{}/g_{pph}^{}$.
There are other restrictions that we need to take into account.

As in the SM+D, for \,$m_D^{}<m_h^{}/2$\, the invisible channel \,$h\to DD$\, is open
and has a rate given by Eq.\,(\ref{Gh2DD}), with $\lambda$ being replaced by $\lambda_h$.
The branching fraction of \,$h\to DD$\, must then be consistent with the LHC measurement
on the Higgs invisible decay, and so for this darkon mass range we again impose
the bound in Eq.\,(\ref{lhclimit}).

Since the extra Higgs particles in the THDM exist due to the second doublet being present,
they generally affect the so-called oblique electroweak parameters $S$ and $T$ which encode
the impact of new physics coupled the standard SU(2)$_L$ gauge boson~\cite{Peskin:1991sw}.
Thus the new scalars must also comply with the experimental constraints on these quantities.
To ensure this, we employ the pertinent formulas from Ref.\,\cite{Grimus:2008nb} and
the $S$ and $T$ data from Ref.\,\cite{pdg}.

Lastly, the parameters of the scalar potential \,${\cal V}={\cal V}_D+{\cal V}_H$\,
in Eq.\,(\ref{pot}) need to fulfill a\,\,number of theoretical conditions.
The quartic couplings in $\cal V$ cannot be too big individually, for otherwise
the theory will no longer be perturbative.
Another requirement is that $\cal V$ must be stable, implying that it has to be bounded
from below to prevent it from becoming infinitely negative for arbitrarily large fields.
It is also essential to ensure that the (tree level) amplitudes for scalar-scalar
scattering at high energies do not violate unitarity constraints.
We address these conditions in more detail in Appendix\,\,\ref{theory-reqs}.
They can be consequential in restraining parts of the model parameter space, especially
for $m_D^{}$ less than ${\cal O}$(100\,GeV), as some of our examples will later demonstrate.

\begin{table}[t] \small
\begin{tabular}{|c|cccccc|cccccccc|} \hline \scriptsize\,Set\, &
$\alpha$ & $\beta$ & $\displaystyle\frac{m_H^{}}{\scriptstyle\rm GeV}$ &
$\displaystyle\frac{m_A^{}}{\scriptstyle\rm GeV}$ &
$\displaystyle\frac{m_{H^\pm}^{}}{\scriptstyle\rm GeV}$ &
$\displaystyle\frac{m_{12}^2}{\scriptstyle\rm GeV^2}$ & $k_V^h$ & $k_u^h$ &
$\displaystyle\frac{k_d^h}{k_u^h}$ & $k_V^H$ & $k_u^H$ & $k_d^H$ &
$\displaystyle\frac{g_{pph}^{}}{10^{-5}}$
& $\displaystyle\frac{f_n^{}}{f_p}\vphantom{|_{\int_\int^\int}^{\int_\int^\int}}$
\\ \hline\hline
1 & ~0.117~$\vphantom{\int^|}$ & $1.428$ & ~470~ & 500 & 550 & \,31000\, & ~0.966~ & 1.003
& ~$-$0.818~ & 0.257 & 0.118 & ~6.98~ & ~10.6~ & \,+0.658\, \\
2 & ~0.141~$\vphantom{\int^|}$ & $1.422$ & ~550~ & 520 & 540 &   44000   & ~0.958~ & 1.001
& ~$-$0.947~ & 0.286 & 0.142 & 6.68 & ~3.29~ &   $-$0.197   \\
3 & ~0.206~$\vphantom{\int^|}$ & $1.357$ & ~515~ & 560 & 570 &   55000   & ~0.913~ & 1.002
& ~$-$0.962~ & ~0.408~ & 0.209 & 4.61 & ~2.42~ & $-$0.646   \\
\hline \end{tabular}
\caption{Sample values of input parameters $\alpha$, $\beta$, $m_{H,A,H^\pm}$, and $m_{12}^2$
in the \,$\lambda_H=0$\, scenario and the resulting values of several quantities,
including \,$f_n^{}/f_p^{}=g_{nnh}^{}/g_{pph}^{}$.\label{lambdaH=0}}
\end{table}
\begin{figure}[b]
\includegraphics[width=8cm]{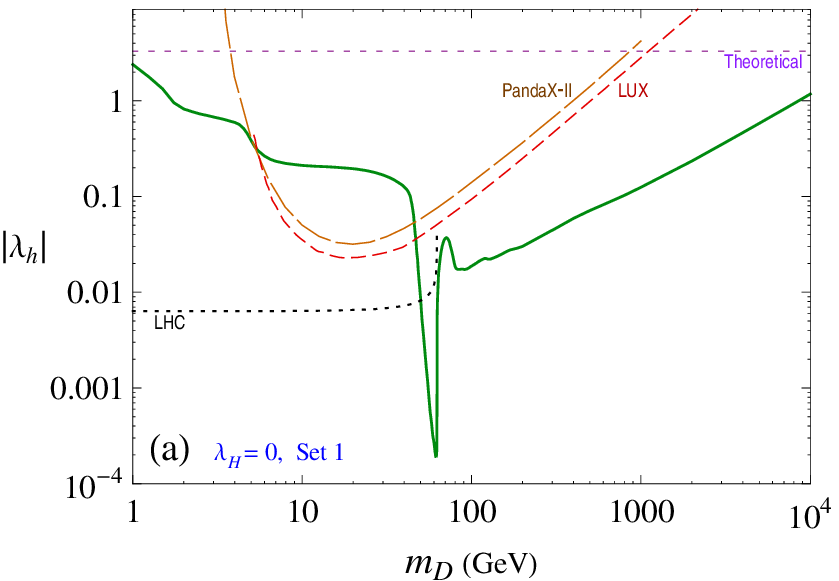} ~ ~
\includegraphics[width=8cm]{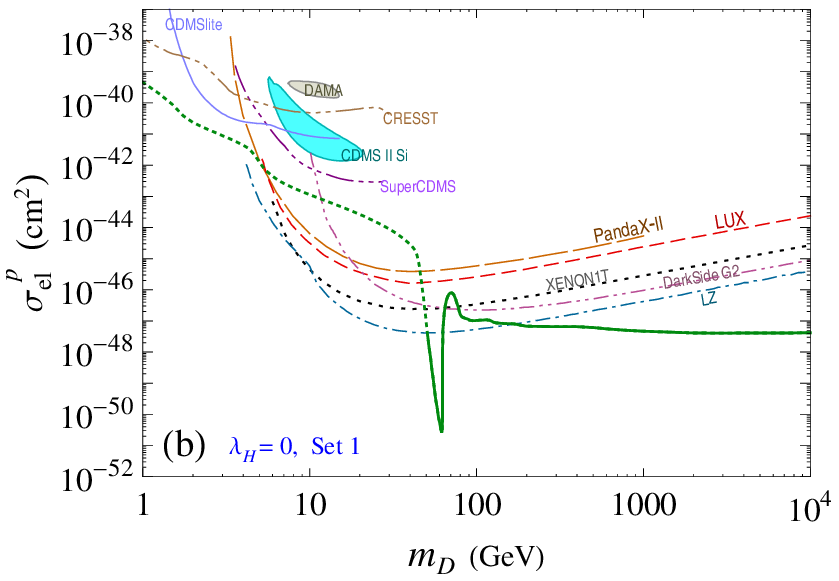}\vspace{-5pt}
\caption{(a) The darkon-$h$ coupling $\lambda_h$ consistent with the relic data
(green curve) versus darkon mass in the THDM\,II+D with \,$\lambda_H=0$\, and
input numbers from Set\,\,1 in Table\,\,\ref{lambdaH=0}.
Also plotted are upper limits from the theoretical conditions mentioned in the text
(horizontal purple dotted-line), the LHC Higgs invisible decay data (black dotted-curve),
and the latest LUX (red dashed-curve) and PandaX-II (orange dashed-curve) results.
(b) The corresponding darkon-proton cross-section $\sigma_{\rm el}^p$ (green curve),
compared to its counterparts translated from the $\sigma_{\rm el}^N$ data and projections
in Fig.\,\ref{expt-plots}(a) using Eq.\,(\ref{ivdm}) with $f_n^{}/f_p^{}$ from Set\,\,1 in
Table\,\,\ref{lambdaH=0}.
The dotted portion of the green curve is excluded by the LHC bound
in (a).\label{2hdm+d-lh-plots1}}
\end{figure}

To illustrate the viable parameter space in this scenario, in the second to seventh columns
of Table\,\,\ref{lambdaH=0} we put together a few sample sets of input parameters which are
consistent with Eq.\,(\ref{khconstr}) and the requirements described in the last two paragraphs.
The eighth to twelfth columns contain the resulting values of several quantities.
With the input numbers from Set\,\,1 in the table, we show in Fig.\,\,\ref{2hdm+d-lh-plots1}(a)
the $\lambda_h$ region evaluated from the observed relic density.
We also display the upper limits on $\lambda_h$ inferred from Eq.\,(\ref{lhclimit}) for
the \,$h\to DD$\, limit (black dotted curve), from the latest LUX~\cite{lux} and
PandaX-II~\cite{pandax} searches, and from the aforementioned theoretical demands for
perturbativity, potential stability, and unitarity.

The plot in Fig.\,\,\ref{2hdm+d-lh-plots1}(b) exhibits the corresponding prediction for
$\sigma_{\rm el}^p$ (green curve) compared to its empirical counterparts obtained from
the data depicted in Fig.\,\ref{expt-plots}(a) by employing Eq.\,(\ref{ivdm})
with \,$f_n^{}/f_p^{}=0.658$\, from Set\,\,1 in Table\,\,\ref{lambdaH=0}.
One observes that the \,$m_D^{}<50$\,GeV\, region, represented by the dotted section of
the green curve, is incompatible with the LHC constraint on \,$h\to DD$\, and
a portion of it is also excluded by LUX and PandaX-II.
The green solid curve is below all of the existing limits from direct searches and for
a narrow range of $m_D^{}$ lies not far under the LUX line.
Upcoming quests with XENON1T as well as DarkSide G2 will apparently be sensitive to only
a small section of the green solid curve, below 100 GeV, whereas LZ can expectedly reach
more of it, from about 63 to 170 GeV.

\begin{figure}[b]
\includegraphics[width=8cm]{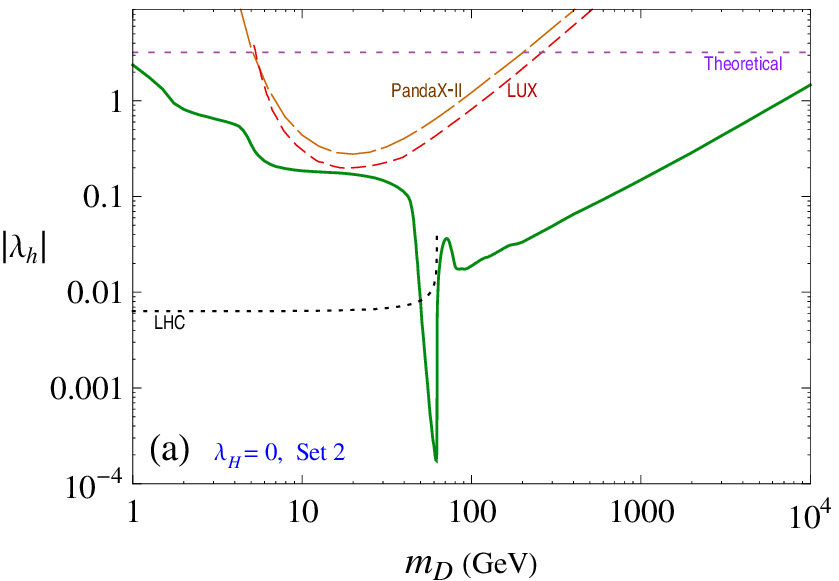} ~ ~
\includegraphics[width=8cm]{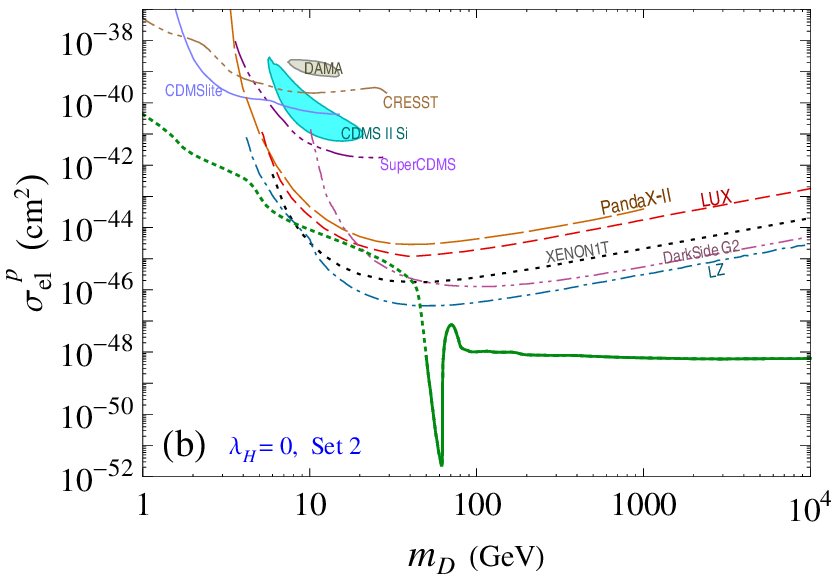}\vspace{-7pt}\\
\includegraphics[width=8cm]{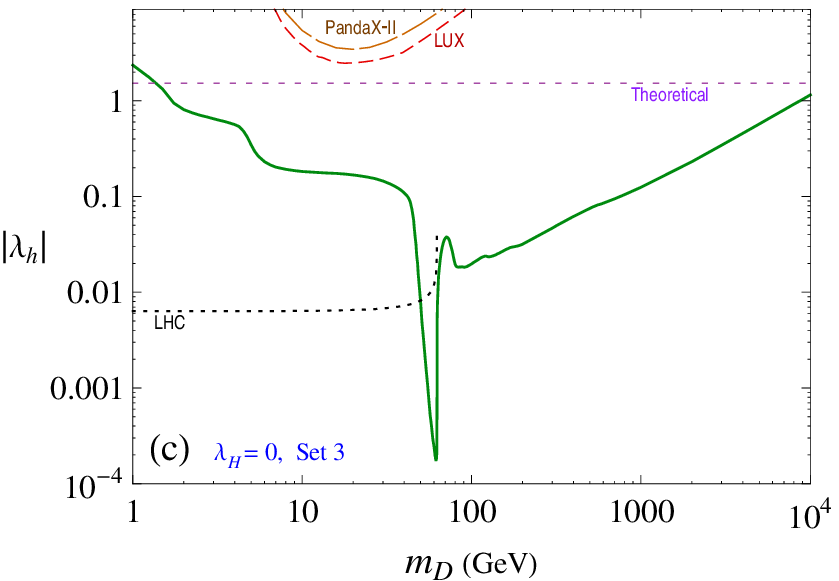} ~ ~
\includegraphics[width=8cm]{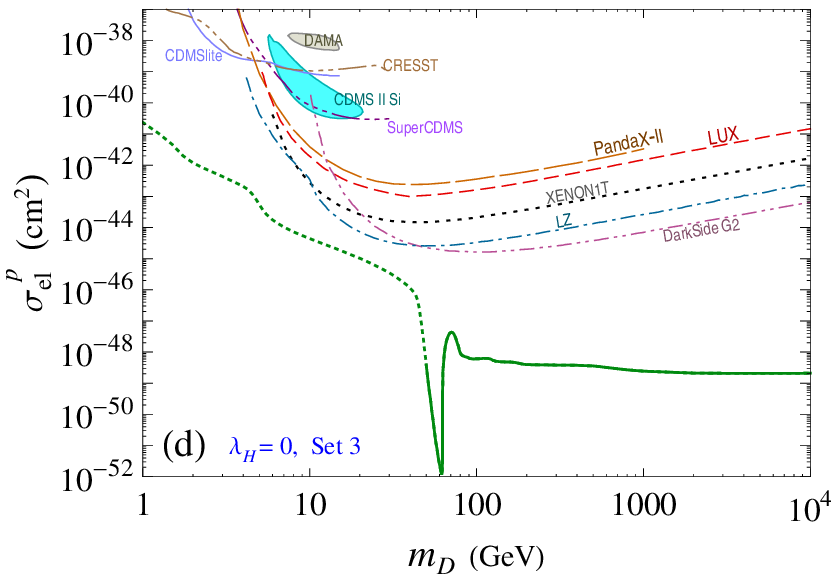}\vspace{-5pt}
\caption{The same as Fig.\,\ref{2hdm+d-lh-plots1}, except the input parameters are
from Set\,\,2 (a,b) and Set 3 (c,d) in Table\,\ref{lambdaH=0}.\label{2hdm+d-lh-plots2}}
\end{figure}

For further illustrations, in Fig.\,\,\ref{2hdm+d-lh-plots2} we graph analogous results
with the input numbers from Sets 2 and 3 in Table\,\,\ref{lambdaH=0}.
Their $f_n^{}/f_p^{}$ values are lower than that in Set 1, making the darkon more xenophobic
and therefore harder to discover with xenon-based detectors, as can also be
deduced from Fig.\,\ref{sigmaratio}.
Especially, in these instances the predictions for $\sigma_{\rm el}^p$ (green solid curves)
are far less than the available experimental bounds and may be out of reach for
direct searches in the not-too-distant future.

\begin{figure}[t]
\includegraphics[width=8cm]{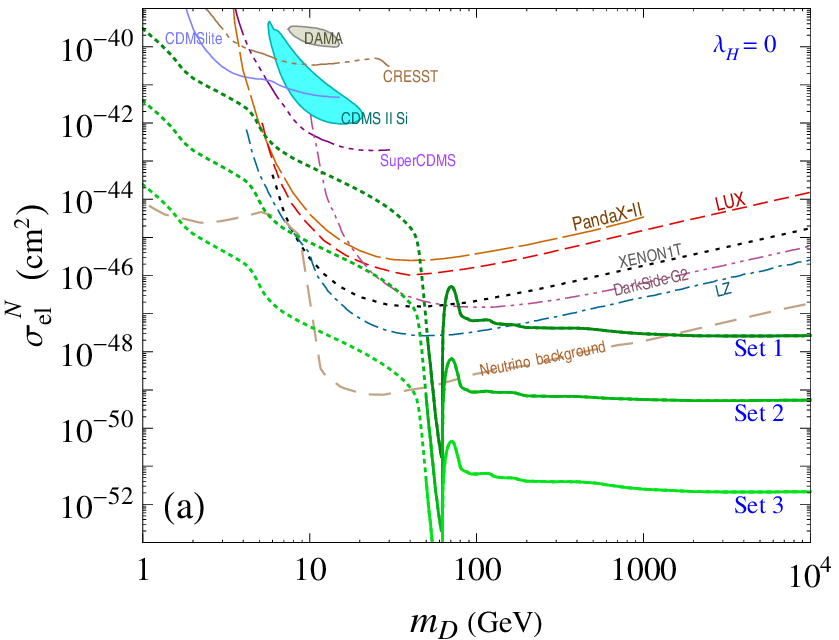}%
\includegraphics[width=8cm]{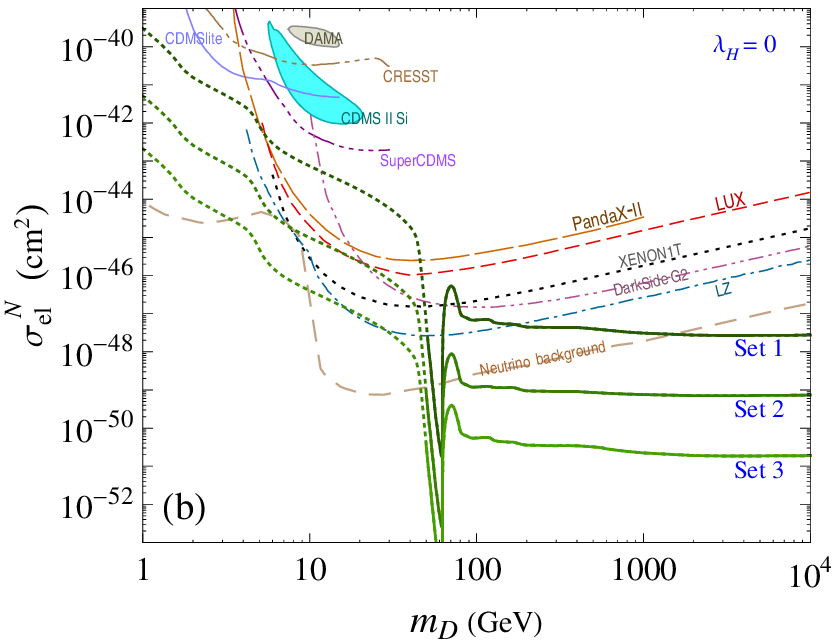}\vspace{-1ex}
\caption{The predictions for darkon-nucleon cross-section $\sigma_{\rm el}^N$ (green curves)
corresponding to Sets 1-3 in Table\,\,\ref{lambdaH=0} for (a) xenon and (b) argon targets
as described in the text, compared to the same data and projections as in
Fig.\,\,\ref{expt-plots}(a).
The dotted portions of the green curves are excluded as in
Figs.\,\,\ref{2hdm+d-lh-plots1} and\,\,\ref{2hdm+d-lh-plots2}.\label{2hdm+d-lh-sigmaelN}}
\end{figure}

For a more straightforward comparison between the model predictions and direct search results,
which are typically reported in terms of the DM-nucleon cross-section $\sigma_{\rm el}^N$,
we have converted the calculated $\sigma_{\rm el}^p$ in Figs.\,\,\ref{2hdm+d-lh-plots1}
and\,\,\ref{2hdm+d-lh-plots2} to the three (green) $\sigma_{\rm el}^N$ curves in
Fig.\,\,\ref{2hdm+d-lh-sigmaelN}(a) using Eq.\,(\ref{ivdm}) with the $f_n^{}/f_p^{}$ values
from Table\,\,\ref{lambdaH=0} and assuming that the target material in the detector is xenon.
Recalling that the DarkSide G2 experiment will employ an argon target~\cite{dsg2}, we plot
the corresponding predictions for $\sigma_{\rm el}^N$ assuming an argon target instead in
Fig.\,\,\ref{2hdm+d-lh-sigmaelN}(b), which reveals some visible differences from
Fig.\,\,\ref{2hdm+d-lh-sigmaelN}(a) in the predictions with \,$f_n^{}/f_p^{}<0$,\,
as Fig.\,\ref{sigmaratio} would imply as well.
Also shown are the same data and projections as in Fig.\,\,\ref{expt-plots}(a).
From Fig.\,\,\ref{2hdm+d-lh-sigmaelN}, we can conclude that near-future direct detection
experiments will be sensitive to only a rather limited part of the $h$-portal THDM\,II+D
parameter space.
We notice specifically that the predicted $\sigma_{\rm el}^N$ in much of
the \,$m_D^{}>100$\,GeV\, region is under the neutrino-background floor.

\subsection{\boldmath$\lambda_h=0$}

In this scenario, the cross section of the darkon annihilation into THDM particles is
\begin{eqnarray}
\sigma_{\rm ann}^{} \,=\, \sigma(DD\to H^*\to X_{\textsc{sm}}) \,+\,
\raisebox{-7pt}{\Large$\stackrel{\sum}{\mbox{\scriptsize${\texttt s}_1{\texttt s}_2$}}$}\,
\sigma(DD\to{\texttt s}_1{\texttt s}_2) \,,
\end{eqnarray}
where
\begin{eqnarray}
\sigma(DD\to H^*\to X_{\textsc{sm}}) \,=\,
\frac{4\lambda_H^2v^2}{\big(m_H^2-s\big)\raisebox{1pt}{$^2$}+\Gamma_H^2m_H^2}~
\frac{\sum_i\Gamma\big(\tilde H\to X_{i,\textsc{sm}}\big)}{\sqrt{s-4m_D^2}} \,,
\end{eqnarray}
with $\tilde H$ being a virtual $H$ having the same couplings as the physical $H$ and
an invariant mass equal to $\sqrt s$, and the sum in $\sigma_{\rm ann}^{}$ is again over
\,${\texttt s}_1{\texttt s}_2=hh,hH,HH,AA,H^+H^-$.\,
For the $H$-mediated darkon-proton scattering, \,$Dp\to Dp$,\, the cross section is
\begin{eqnarray} \label{Dp2Dp'}
\sigma_{\rm el}^p \,=\,
\frac{\lambda_H^2\,g_{ppH}^2\,m_p^2\,v^2}{\pi\,\bigl(m_D^{}+m_p^{}\bigr)^2m_H^4} \,.
\end{eqnarray}
In applying Eq.\,(\ref{ivdm}), we set \,$f_n^{}/f_p^{}=g_{nnH}^{}/g_{ppH}^{}$.\,

\begin{table}[t] \footnotesize
\begin{tabular}{|c|cccccc|cccccccc|} \hline \scriptsize\,Set\, &
$\alpha$ & $\beta$ & $\displaystyle\frac{m_H^{}}{\scriptstyle\rm GeV}$ &
$\displaystyle\frac{m_A^{}}{\scriptstyle\rm GeV}$ &
$\displaystyle\frac{m_{H^\pm}^{}}{\scriptstyle\rm GeV}$ &
$\displaystyle\frac{m_{12}^2}{\scriptstyle\rm GeV^2}$ & $k_V^h$ & $k_u^h$ & $k_d^h$ &
$k_V^H$ & $k_u^H$ & $\displaystyle\frac{k_d^H}{k_u^H}$ &
$\displaystyle\frac{g_{ppH}^{}}{10^{-5}}$
& $\displaystyle\frac{f_n^{}}{f_p}\vphantom{|_{\int_\int^\int}^{\int_\int^\int}}$
\\ \hline\hline
1 & ~$\vphantom{\int^|}-$0.785~ & $0.738$ & ~550~ & 600 & 650 & \,70000\, &
~0.999~ & 1.051 & ~0.955~ & ~0.048~ & $-$1.051 & ~$-$0.910~ & ~$-$5.62~ & \,+0.281\, \\
2 & ~$\vphantom{\int^|}-$0.749~ & $0.723$ & ~610~ & 750 & 760 &   91000 &
~0.995~ & 1.107 & ~0.908~ & ~0.099~ & $-$1.029 & ~$-$0.949~ & ~$-$3.26~ & $-$0.245 \\
3 & ~$\vphantom{\int^|}-$0.676~ & $0.658$ & ~590~ & 610 & 640 &   60000 &
~0.972~ & 1.276 & ~0.791~ & ~0.235~ & $-$1.023 & ~$-$0.964~ & ~$-$2.40~ & $-$0.693 \\
\hline \end{tabular}
\caption{Samples values of input parameters $\alpha$, $\beta$, $m_{H,A,H^\pm}$, and $m_{12}^2$
in the \,$\lambda_h=0$\, scenario and the resulting values of several quantities,
including \,$f_n^{}/f_p^{}=g_{nnH}^{}/g_{ppH}^{}$.\label{lambdah=0}}
\end{table}
\begin{figure}[t]
\includegraphics[width=79mm]{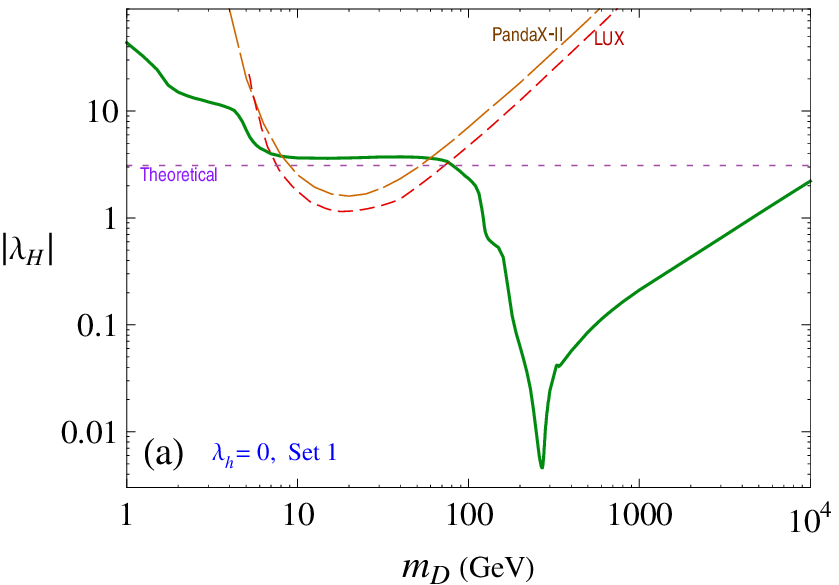} ~ ~
\includegraphics[width=80mm]{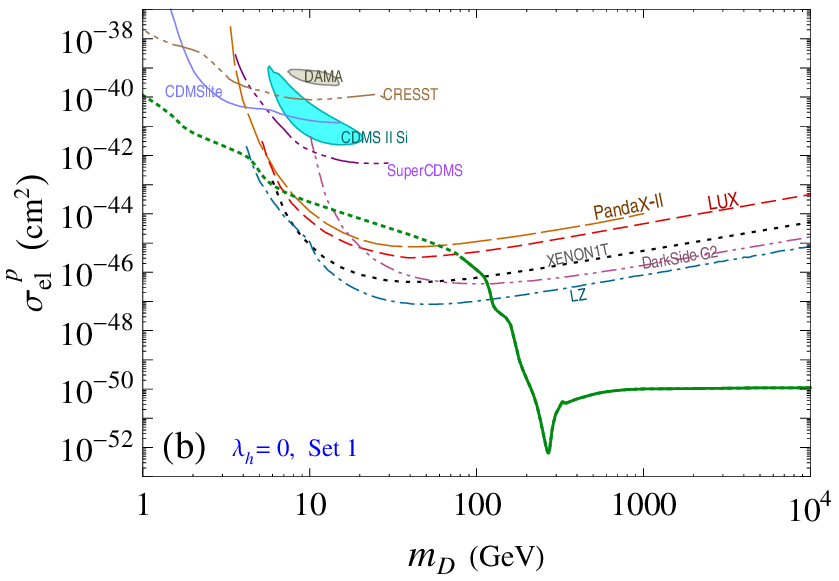}\vspace{-5pt}
\caption{(a) The darkon-$H$ coupling $\lambda_H$ versus darkon mass
in the THDM\,II+D with \,$\lambda_h=0$\, and input numbers from Set\,\,1 in
Table\,\,\ref{lambdah=0}.
(b)~The corresponding darkon-proton cross-section $\sigma_{\rm el}^p$ (green curve),
compared to its counterparts translated from Fig.\,\ref{expt-plots}(a) using
Eq.\,(\ref{ivdm}) with $f_n^{}/f_p^{}$ from Set\,\,1 in Table\,\,\ref{lambdah=0}.
The dotted portion of the green curve is excluded by the theoretical bound
in (a).\label{2hdm+d-lH-plots1}\medskip}
\end{figure}

Similarly to the \,$\lambda_H=0$\, case, here we present three examples, with their
respective sets of input numbers being collected in Table\,\,\ref{lambdah=0}.
We also impose the requirements described earlier in this section, except that the LHC
information on the decay mode \,$h\to\rm invisible$\, is not useful for bounding $\lambda_H$.
Nevertheless, the theoretical conditions for perturbativity, stability of the potential,
and unitarity of high-energy scalar scattering amplitudes turn out to be
consequential in disallowing darkon masses less than 100 GeV.

\begin{figure}[!t]
\includegraphics[width=79mm]{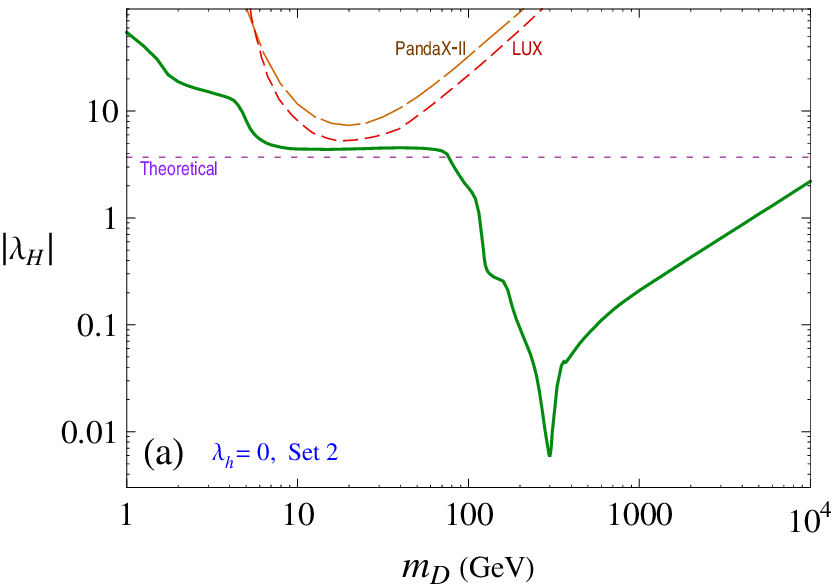} ~ ~
\includegraphics[width=80mm]{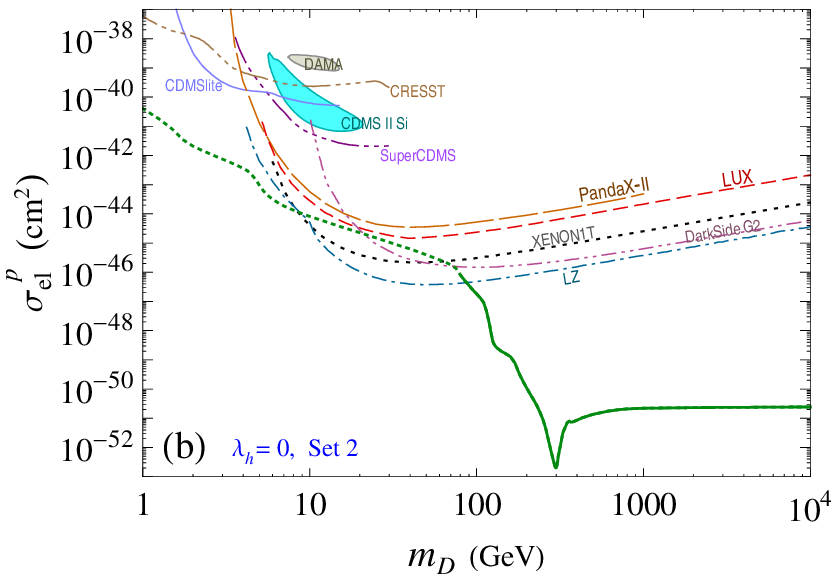}\vspace{-7pt}\\
\includegraphics[width=79mm]{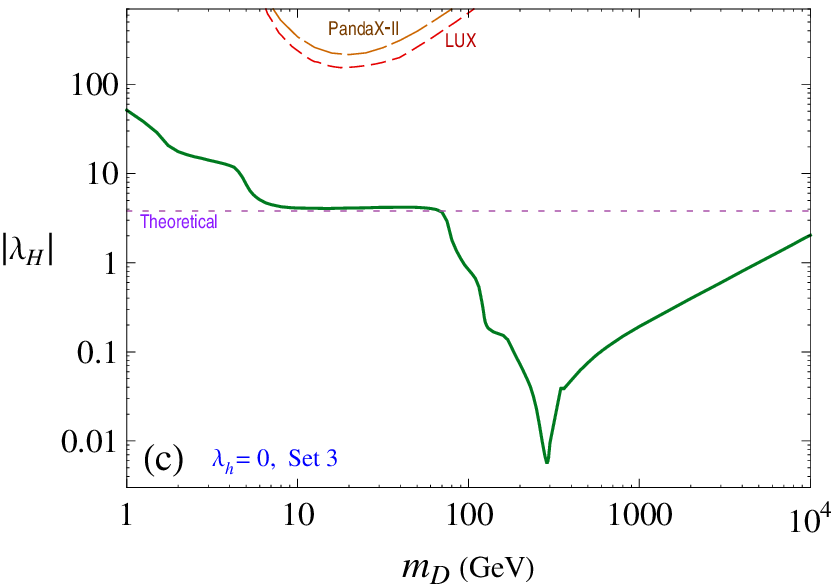} ~ ~
\includegraphics[width=80mm]{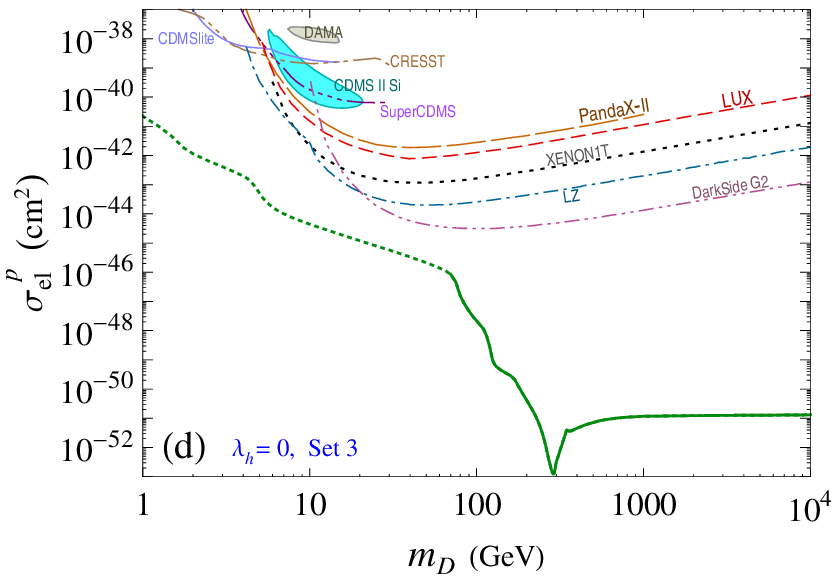}\vspace{-5pt}
\caption{The same as Fig.\,\ref{2hdm+d-lH-plots1}, except the input parameters are
from Set\,\,2 (a,b) and Set 3 (c,d) in Table\,\ref{lambdah=0}.\label{2hdm+d-lH-plots2}}
\end{figure}

The input numbers from Set 1 (Sets 2 and 3) in Table\,\,\ref{lambdah=0} lead to the graphs
in Fig.\,\,\ref{2hdm+d-lH-plots1} (\ref{2hdm+d-lH-plots2}).
In these figures, we see that the $|\lambda_H|$ values extracted from the relic density data
tend to be bigger than their $\lambda_h$ counterparts in the \,$\lambda_H=0$\, instances.
This is because the $H$-mediated annihilation rate is relatively more suppressed due to
\,$m_H^{}>m_h^{}$.\,
As a consequence, more of the low-$m_D^{}$ regions are in conflict with the restrictions
from the aforementioned theoretical requirements.
Furthermore, in Fig.\,\,\ref{2hdm+d-lH-plots1}(b), like in
Fig.\,\,\ref{2hdm+d-lh-plots1}(b), there is a small range of the solid green curve, around
its leftmost end, that is close to the LUX and PandaX-II limits.

\begin{figure}[!t] \smallskip
\includegraphics[width=8cm]{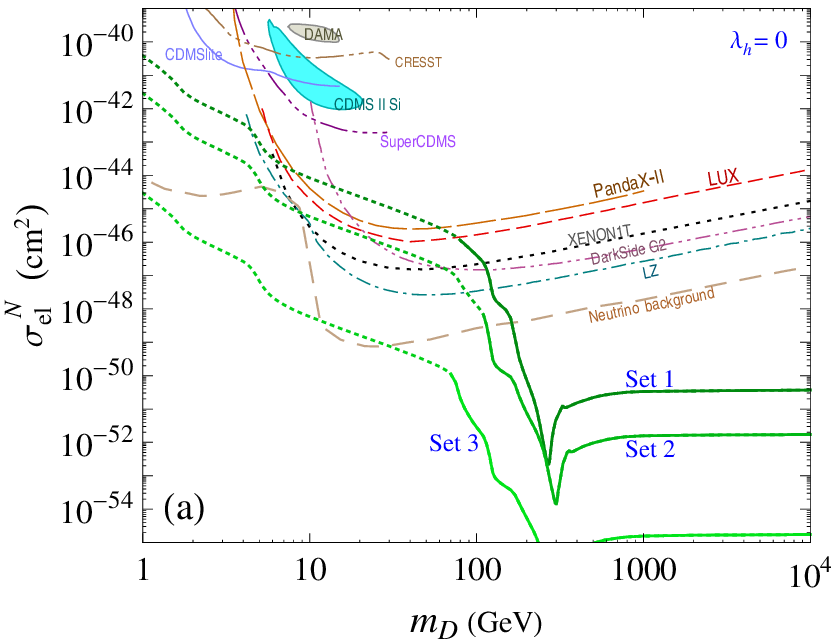}%
\includegraphics[width=8cm]{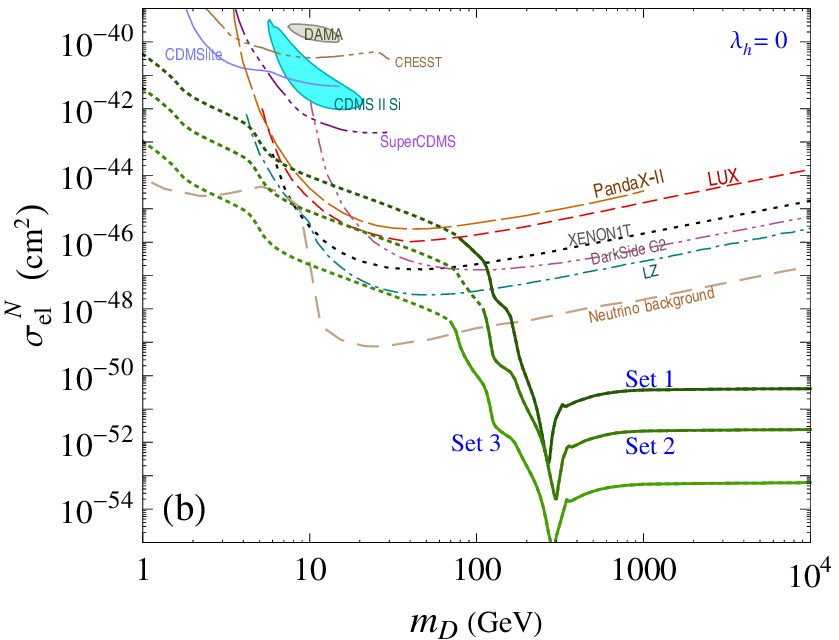}\vspace{-1ex}
\caption{The predictions for darkon-nucleon cross-section $\sigma_{\rm el}^N$ (green curves)
corresponding to Sets 1-3 in Table\,\,\ref{lambdah=0} for (a) xenon and (b) argon targets as
described in the text, compared to the same data and projections as in Fig.\,\,\ref{expt-plots}(a).
The dotted portions of the green curves are excluded as in
Figs.\,\,\ref{2hdm+d-lH-plots1} and\,\,\ref{2hdm+d-lH-plots2}.\label{2hdm+d-lH-sigmaelN}}
\end{figure}

As in the previous subsection, assuming xenon to be the target material, we have translated
the predicted $\sigma_{\rm el}^p$ in Figs.\,\,\ref{2hdm+d-lH-plots1}
and\,\,\ref{2hdm+d-lH-plots2} into the three (green)
$\sigma_{\rm el}^N$ curves in Fig.\,\,\ref{2hdm+d-lH-sigmaelN}(a) in order to provide
a more direct comparison with experimental results.
If the target is argon instead and \,$f_n^{}/f_p^{}<0$,\, the $\sigma_{\rm el}^N$ predictions
can be visibly greater, as depicted in Fig.\,\,\ref{2hdm+d-lH-sigmaelN}(b).

For darkon masses above 100 GeV, the majority of the $\sigma_{\rm el}^N$ predictions
in Fig.\,\,\ref{2hdm+d-lH-sigmaelN} appear to lie under the neutrino-background floor,
more in these instances than those in the \,$\lambda_H=0$\, case.
Thus, the \,$\lambda_h=0$\, scenario is likely to be comparatively more challenging to
probe with direct searches.


\section{Conclusions\label{conclusion}}

We have explored some of the implications of the most recent null results of WIMP DM direct
searches by LUX and PandaX-II.
For Higgs-portal scalar WIMP DM models, the new limits have eliminated any possibility to
accommodate low-mass DM undergoing spin-independent elastic scattering off nucleons that was
suggested by the potentially positive results of the DAMA and CDMS II Si experiments,
even after invoking the mechanism of isospin violation in DM-nucleon interactions.
We have studied particularly how the LUX and PandaX-II results probe the parameter space
of the simplest Higgs-portal scalar DM models, namely the SM+D, which is the SM
plus a\,\,real scalar singlet called darkon, and the THDM\,II+D,
which is the two-Higgs-doublet model of type II combined with a darkon.
In the THDM\,II+D we entertain the possibility that the 125-GeV Higgs boson, $h$, is
the lightest one of the physical members of the scalar doublets.
Our analysis takes into account various constraints from LHC data on the Yukawa
\mbox{couplings of $h$}, its couplings to gauge bosons, and its invisible decay mode.
Also pertinent are restrictions from oblique electroweak precision measurements and
from theoretical considerations regarding perturbativity, vacuum stability, and unitarity.
In the SM+D case, $h$ is the only portal between the DM and SM sectors, while in
the THDM\,II+D one or both of the $CP$-even Higgs bosons, \mbox{$h$ and the heavier $H$},
can be the portals.

We find that in scenarios with $h$ being the only portal the LHC information on
\,$h\to\rm invisible$\, places a significant restraint on the darkon-$h$ coupling
and rules out the \,$m_D^{}\le m_h^{}/2$\, region, except a small range near the resonance
point \,$m_D^{}=m_h^{}/2$.\,
We also find that for \,$m_D^{}>m_h^{}/2$\, in the SM+D the new LUX and PandaX-II limits
exclude masses up to 450 GeV or so, but in the $h$-portal THDM\,II+D they can be recovered
due to suppression of the Higgs-nucleon coupling,\,\,$g_{{\cal NN}h}^{}$, at some values of
the product \,$\tan\alpha\,\tan\beta$.\,
In contrast, in the THDM\,II+D scenario with $H$ being the sole portal, the
\,$h\to\rm invisible$\, bound does not apply to the much heavier $H$, and the LUX and
PandaX-II limits can be evaded due to suppression of $g_{{\cal NN}H}^{}$ at some
values of \,$\cot\alpha\,\tan\beta$.\,
However, in this case our examples demonstrate that the foregoing theoretical requirements
are consequential and disallow most of the \,$m_D^{}<100$\,GeV\, region.
Thus, darkon masses below \,$m_D^{}\simeq50$\,GeV\, are ruled out in the SM+D by LHC data
and very likely so in the THDM\,II+D by the LHC and theoretical restrictions.
For higher masses, lower parts of the dip around \,$m_D^{}=m_h^{}/2$\, in the $h$-portal cases
will remain viable for the foreseeable future, and beyond the $h$-resonance area the region
up to roughly 3.5, 10, and 20\,\,TeV in the SM+D will be testable by XENON1T, DarkSide G2,
and LZ, respectively.
For \,$m_D^{}>100$\,GeV\, in the THDM\,II+D there is generally ample parameter space that
yields a\,\,darkon-nucleon cross-section below the neutrino-background floor and is therefore
likely to elude direct detection experiments in the future which lack directional sensitivity.
Finally, we point out that the considerable suppression of $g_{{\cal NN}\cal H}^{}$ is
accompanied by $g_{pp\cal H}^{}$ and $g_{nn\cal H}^{}$ manifesting sizable isospin breaking,
as illustrated in our examples.

\acknowledgments

This research was supported in part by MOE Academic Excellence Program (Grant No. 102R891505)
and NCTS of ROC.
X.-G. He was also supported in part by MOST of ROC (Grant No.~MOST104-2112-M-002-015-MY3)
and in part by NSFC (Grant Nos.~11175115 and 11575111), Key Laboratory for Particle Physics,
Astrophysics and Cosmology, Ministry of Education, and Shanghai Key Laboratory for Particle
Physics and Cosmology (SKLPPC) (Grant No.~11DZ2260700) of PRC.

\appendix

\section{Extra formulas for darkon reactions\label{formulas}}

To extract the darkon-Higgs coupling which enters the annihilation cross-section
$\sigma_{\rm ann}$, we employ its thermal average~\cite{Gondolo:1990dk}
\begin{eqnarray} \label{sv}
\langle\sigma v_{\rm rel}^{}\rangle \,=\, \frac{x}{8 m_D^{5\;}K_2^2(x)}
\int_{4m_D^2}^\infty ds\;\sqrt s\,\bigl(s-4m_D^2\bigr)\;
K_1^{}\bigl(\sqrt s\,x/m_D^{}\bigr)\,\sigma_{\rm ann}^{} \,,
\end{eqnarray}
where $v_{\rm rel}$ is the relative speed of the DM pair, $K_r$ is the modified Bessel
function of the second kind of order $r$ and $x$ can be set to its freeze out value
\,$x=x_f^{}$,\, which is related to $\langle\sigma v_{\rm rel}^{}\rangle$ by~\cite{Kolb:1990vq}
\begin{eqnarray} \label{xf}
x_f^{} \,=\,
\ln\frac{0.038\,m_D^{}\,m_{\rm Pl}^{}\,\langle\sigma v_{\rm rel}^{}\rangle}{\sqrt{g_*^{}x_f^{}}} \,,
\end{eqnarray}
with \,$m_{\rm Pl}=1.22 \times 10^{19}$\, GeV being the Planck mass and $g_*^{}$ is the total
number of effectively relativistic degrees of freedom below the freeze-out
temperature \,$T_f^{}=m_D^{}/x_f^{}$.\,
In addition, we adopt the numerical values of $\langle\sigma v_{\rm rel}^{}\rangle$ versus $m_D^{}$
determined in Ref.~\cite{Steigman:2012nb}, as well as the latest relic density data
\,$\Omega\hat h^2=0.1197\pm0.0022$\, \cite{planck}, with $\hat h$ being the Hubble parameter.

\begin{figure}[t]
\includegraphics[width=10cm]{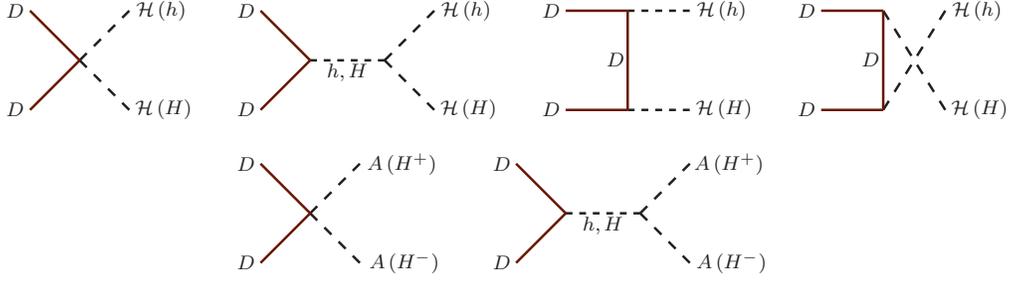}\vspace{-1ex}
\caption{Feynman diagrams contributing to \,$DD\to hh,HH,hH,AA,H^+H^-$.\,\label{dd2hh}}
\end{figure}

In the THDM\,II+D, if kinematically allowed, a darkon pair can annihilate into a pair of Higgs
bosons, \,$DD\to hh,hH,HH,AA,H^+H^-$,\, induced by the diagrams drawn in Fig.\,\ref{dd2hh}.
They lead to the cross sections
\begin{eqnarray} \label{DD->hh}
\sigma(DD\to{\cal HH}) &=&
\frac{\lambda_{\cal H}^2 v^2}{\beta_D^2\pi s^2} \Bigg( \frac{{\cal M}_{\cal HH}}{2}
+ \frac{2\lambda_{\cal H}^2\,v^2}{2m_{\cal H}^2-s} \Bigg)
\ln\!\left|\frac{s-2m_{\cal H}^2-\beta_D^{}\beta_{\cal H}^{}s}
{s-2m_{\cal H}^2+\beta_D^{}\beta_{\cal H}^{}s}\right|
\nonumber \\ && \! +\;
\frac{\beta_{\cal H}^{}}{\beta_D^{}} \Bigg[ \frac{{\cal M}_{\cal HH}^2}{32\pi s}
+ \frac{\lambda_{\cal H}^4\,v^4}{\pi \big(m_{\cal H}^4+\beta_{\cal H}^2 m_D^2 s\big)s} \Bigg] \,,
\end{eqnarray}
\begin{eqnarray} \label{DD->hH}
\sigma(DD\to hH) &=&
\frac{\lambda_h^{}\lambda_H^{}\,v^2}{\beta_D^2 \pi s^2} \Bigg( {\cal M}_{hH}^{}
+ \frac{4\lambda_h^{}\lambda_H^{}\,v^2}{m_h^2+m_H^2-s} \Bigg)
\ln\!\left|\frac{s-m_h^2-m_H^2 - \beta_D^{}\,{\cal K}^{\frac{1}{2}}\big(s,m_h^2,m_H^2\big)}
{s-m_h^2-m_H^2 + \beta_D^{}\,{\cal K}^{\frac{1}{2}}\big(s,m_h^2,m_H^2\big)}\right|
\nonumber \\ && \! +\;
\frac{{\cal K}^{\frac{1}{2}}\big(s,m_h^2,m_H^2\big)}{\beta_D^{} \pi s}
\Bigg( \frac{{\cal M}_{hH}^2}{16 s}
+ \frac{2\lambda_h^2\lambda_H^2\,v^4}{m_h^2m_H^2s+{\cal K}\big(s,m_h^2,m_H^2\big)m_D^2}
\Bigg) \,,
\end{eqnarray}
\begin{eqnarray} \label{DD->AA}
\sigma(DD\to AA) \,=\, \frac{\beta_A^{}\,{\cal M}_{AA}^2}{32\beta_D^{}\pi s}\,, ~~~~ ~~~
\sigma(DD\to H^+H^-) \,=\, \frac{\beta_{H^\pm}^{}\,{\cal M}_{H^+H^-}^2}{16\beta_D^{}\pi s} \,, ~~~
\end{eqnarray}
where $\sqrt s$ is the c.m. energy of the darkon pair, \,${\cal HH}=hh,HH$,\,
\begin{eqnarray} \label{MXY}
\beta_{\texttt X}^{} &=& \sqrt{1-\frac{4m_{\texttt X}^2}{s}} \,, ~~~~ ~~~
{\mathcal K}(x,y,z) \,=\, x^2+y^2+z^2-2(x y+y z+x z) \,,
\\
{\cal M}_{\texttt{XY}}^{} &=& 2\lambda_{\texttt{XY}}^{} \,+\,
\frac{2\lambda_h^{}\lambda_{h\texttt{XY}}^{}\,\big(s-m_h^2\big)v^2}
{\big(s-m_h^2\big)\raisebox{1pt}{$^2$}+\Gamma_h^2m_h^2}
\,+\, \frac{2\lambda_H^{}\lambda_{H\texttt{XY}}^{}\,\big(s-m_H^2\big)v^2}
{\big(s-m_H^2\big)\raisebox{1pt}{$^2$}+\Gamma_H^2m_H^2} \,, ~~~~~ ~~
\nonumber \\ \texttt{XY} &=& hh,hH,HH,AA,H^+H^- \,,
\end{eqnarray}
with $\lambda_{\texttt{XY}}$ being given by Eq.\,\,(\ref{lambdah}) and
\begin{eqnarray} \label{lambdas}
\lambda_{hhh}^{} &=&
\frac{c_\alpha^3c_\beta^{}-s_\alpha^3s_\beta^{}}{c_\beta^{}s_\beta^{}}~\frac{3m_h^2}{v^2} \,-\,
\frac{3c_{\alpha+\beta}^{}\,c_{\alpha-\beta}^2}{c_\beta^2\,s_\beta^2}
\frac{m_{12}^2}{v^2} \,, ~~~~
\nonumber \\
\lambda_{hhH}^{} &=& \frac{c_{\beta-\alpha}^{}}{c_\beta^{}s_\beta^{}} \Bigg[
s_{2\alpha}^{}\,\frac{2m_h^2+m_H^2}{2v^2} +
\Bigg(1-\frac{3s_{2\alpha}^{}}{s_{2\beta}^{}}\Bigg)\frac{m_{12}^2}{v^2} \Bigg]
\,=\, \lambda_{Hhh}^{} \,,
\nonumber \\
\lambda_{hHH}^{} &=& \frac{s_{\alpha-\beta}^{}}{c_\beta^{}s_\beta^{}} \Bigg[
s_{2\alpha}^{}\,\frac{m_h^2+2m_H^2}{2v^2} \,-\,
\Bigg(1+\frac{3s_{2\alpha}^{}}{s_{2\beta}^{}}\Bigg)\frac{m_{12}^2}{v^2} \Bigg]
\,=\, \lambda_{HhH}^{} \,,
\nonumber \\
\lambda_{hH^+H^-}^{} &=& \frac{c_\alpha^{}c_\beta^3-s_\alpha^{}s_\beta^3}{c_\beta^{}s_\beta^{}}~
\frac{m_h^2}{v^2} +
2s_{\beta-\alpha}^{} \frac{m_{H^\pm}^2}{v^2}
- \frac{c_{\alpha+\beta}^{}}{c_\beta^2s_\beta^2} \frac{m_{12}^2}{v^2}
\,=\, \lambda_{hAA}^{} + 2s_{\beta-\alpha}^{}\frac{m_{H^\pm}^2-m_A^2}{v^2} \,,
\nonumber \\
\lambda_{HHH}^{} &=& \frac{c_\alpha^3s_\beta^{}+s_\alpha^3c_\beta^{}}{c_\beta^{}s_\beta^{}}~
\frac{3m_H^2}{v^2} \,-\,
\frac{3s_{\alpha+\beta}^{}\,s_{\alpha-\beta}^2}{c_\beta^2s_\beta^2}\frac{m_{12}^2}{v^2} \,,
\nonumber \\
\lambda_{HH^+H^-}^{} &=& \frac{c_\alpha^{}s_\beta^3+s_\alpha^{}c_\beta^3}{c_\beta^{}s_\beta^{}}~
\frac{m_H^2}{v^2} +
2c_{\beta-\alpha}^{} \frac{m_{H^\pm}^2}{v^2}
- \frac{s_{\alpha+\beta}^{}}{c_\beta^2s_\beta^2} \frac{m_{12}^2}{v^2}
\,=\, \lambda_{HAA}^{} + 2c_{\beta-\alpha}^{}\frac{m_{H^\pm}^2-m_A^2}{v^2} \,. ~~~
\end{eqnarray}
In Eqs.\,\,(\ref{DD->hh}) and\,\,(\ref{DD->hH}), we have dropped terms with two powers of
$\Gamma_{h,H}$ in the numerators.
In the scenarios we look at, $\Gamma_H$ receives contributions not only from rates of
the fermion and gauge-boson decay modes of $H$, similarly to those of $h$, but also from
\begin{eqnarray}
\Gamma(H\to DD) \,=\, \frac{\lambda_{H\,}^2v^2}{8\pi\,m_H^{}}\sqrt{1-\frac{4m_D^2}{m_H^2}} \,,
~~~ ~~~~
\Gamma(H\to hh) \,=\, \frac{\lambda_{hhH\,}^2v^2}{8\pi\,m_H^{}}\sqrt{1-\frac{4m_h^2}{m_H^2}}
\end{eqnarray}
once these channels are open.
The $\sigma(DD\to hh)$ formula in Eq.\,(\ref{DD->hH}) is applicable to the SM+D, in which case
there is only one coupling for the darkon-Higgs interaction, \,$\lambda_h=\lambda_{hh}=\lambda$,\,
and there is no $H$ contribution, \,$\lambda_H=\lambda_{hhH}=0$.\,

The parameters $f_q^{\cal N}$ in Eq.\,(\ref{gNNH}) depend on the so-called pion-nucleon
sigma term $\sigma_{\pi N}$, which is not yet well-determined.
To minimize the prediction for $\sigma_{\rm el}^{N}$ in view of the stringent experimental
restraints, we estimate $f_q^{\cal N}$ using the results of
Refs.\,\cite{He:2008qm,He:2011gc} with \,$\sigma_{\pi N}=30~$MeV.\,
This yields
\begin{eqnarray}
f_u^p &=& 0.01370 \,, ~~~~~ f_d^p \,=\, 0.01686 \,, ~~~~~ f_s^p \,=\, 0.06305 \,, ~~~~~
f_{c,b,t}^p \,=\, 0.06703 \,,
\nonumber \\
f_u^n &=& 0.00976 \,, ~~~~~ f_d^n \,=\, 0.02359 \,, ~~~~~ f_s^n \,=\, 0.06296 \,, ~~~~~
f_{c,b,t}^n \,=\, 0.06694 \,.
\end{eqnarray}
We note that \,$f_{c,b,t}^{\cal N}\simeq2\big(1-f_u^{\cal N}-f_d^{\cal N}-f_s^{\cal N}\big)/27$.\,

\section{Conditions for perturbativity, vacuum stability, and tree-level
unitarity\label{theory-reqs}}

The parameters of the scalar potential \,${\cal V}={\cal V}_H+{\cal V}_D$\, of the THDM\,II+D
in Eq.\,\,(\ref{pot}) are subject to a number of theoretical constraints.
We adopt the usual assumption that the scalar interactions are in the perturbative regime,
implying that the $\lambda$ parameters in $\cal V$ need to be capped.
Thus, we demand that \,$|\lambda_{1,2,3,4,5}|\le8\pi$,\, like in the THDM scenario without
the darkon~\cite{Kanemura:1999xf}, while for the darkon couplings \,$|\lambda_{D,1D,2D}|\le4\pi$\,
in view of their normalization convention in $\cal V$.
In what follows, we describe additional requirements which may lead to stronger restraints
on these $\lambda$s.

The requisite stability of $\cal V$ implies that it has to be bounded from below.
In other words, its quartic part
\begin{eqnarray}
{\cal V}_4^{} &=&
\tfrac{1}{2}\lambda_{1\,}^{}\bigl(H_1^\dagger H_1^{}\bigr)\raisebox{1pt}{$^2$} +
\tfrac{1}{2}\lambda_{2\,}^{}\bigl(H_2^\dagger H_2^{}\bigr)\raisebox{1pt}{$^2$} +
\lambda_{3\,}^{}H_1^\dagger H_{1\,}^{}H_2^\dagger H_2^{}
+ \lambda_{4\,}^{}H_1^\dagger H_{2\,}^{}H_2^\dagger H_1^{} \,+\, \tfrac{1}{2}\lambda_5^{} \Big[
\big(H_1^\dagger H_2^{}\big)\raisebox{1pt}{$^2$} + {\rm H.c.} \Big]
\nonumber \\ && \!\! +\;
\tfrac{1}{4}\lambda_D^{}\,D^4 \,+\,
\big(\lambda_{1D\,}^{}H_1^\dagger H_1^{}+\lambda_{2D\,}^{}H_2^\dagger H_2^{}\big)D^2
\end{eqnarray}
must stay positive for arbitrarily large values of the fields.
Expressing
\begin{eqnarray}
H_r^\dagger H_r^{} = \eta_r^2 \,, ~~~~ \eta_r^{} \ge 0 \,, ~~~~ r = 1,2 \,,  ~~~~ ~~~
H_1^\dagger H_2^{} = \eta_1^{}\eta_2^{}\,\rho^2e^{i\theta} \,, ~~~~
0 \le \rho^2 \le 1 \,, ~~~~ {\rm Im}_{\,}\theta = 0 \,, ~~~
\end{eqnarray}
we then have
\begin{eqnarray}
{\cal V}_4^{} &=& \tfrac{1}{2} \Big( \eta_1^2 ~~~ \eta_2^2 ~~~ D^2 \Big)\, {\cal M}_4^{} \left(
\begin{array}{c} \eta_1^2 \\ \eta_2^2 \\ D^2 \end{array}\right) ,
\nonumber \\
{\cal M}_4^{} &=& \!\left( \begin{array}{ccc}
\lambda_1^{} & \lambda_3^{}+[\lambda_4+\lambda_5\,{\rm cos}(2\theta)]\rho^2 & \lambda_{1D}^{} \\
\lambda_3^{}+[\lambda_4+\lambda_5\,{\rm cos}(2\theta)]\rho^2 & \lambda_2^{} & \lambda_{2D}^{} \\
\vphantom{|_{\int}^{}} \lambda_{1D}^{} & \lambda_{2D}^{} & \tfrac{1}{2}\lambda_D^{} \end{array}
\right) .
\end{eqnarray}
For any of $\eta_{1,2}^{}$ and $D$ being large, \,${\cal V}_4^{}>0$\, if ${\cal M}_4^{}$ is
strictly copositive~\cite{copositivity}, and this entails
\begin{eqnarray} \label{stability}
\lambda_r^{} &>& 0 \,, ~~~~ \lambda_D^{} \,>\, 0 \,, ~~~~
\lambda_{rD}^{} \,>\, -\sqrt{\tfrac{1}{2}\lambda_r^{}\lambda_D^{}} \,, ~~~~
\lambda_3 + {\rm min}(0,\lambda_4-|\lambda_5|) \,>\, -\sqrt{\lambda_1\lambda_2} \,,
\vphantom{|_{\int_\int^\int}^{}} \nonumber \\
0  &<&
\lambda_{1D}^{}\sqrt{2\lambda_2} \,+\, \lambda_{2D}^{}\sqrt{2\lambda_1}
+ \Big[ \sqrt{\lambda_1^{}\lambda_2^{}} + \lambda_3 +
{\rm min}(0,\lambda_4-|\lambda_5|) \Big] \sqrt{\lambda_D}
\nonumber \\ && \!\! +\;
\sqrt{\Big(\sqrt{2\lambda_1^{}\lambda_D^{}}+2\lambda_{1D}^{}\Big)
\Big(\sqrt{2\lambda_2^{}\lambda_D^{}}+2\lambda_{2D}^{}\Big)
\Big[ \sqrt{\lambda_1\lambda_2}+\lambda_3+{\rm min}(0,\lambda_4-|\lambda_5|)\Big] } \,, ~~~
\end{eqnarray}
where \,$r=1,2$.\,

\newpage

Another important limitation is that the amplitudes for scalar-scalar scattering
\,$s_1^{}s_2^{}\to s_3^{}s_4^{}$\, at high energies respect unitarity.
Similarly to the THDM case~\cite{Branco:2011iw,Kanemura:1993hm,Akeroyd:2000wc}, for the scalar
pair \,$s_m^{}s_n^{}$\, we can work with the nonphysical fields $h_r^\pm$, $h_r^0$, and
$I_r^0$, as well as $D$.
Accordingly, one can take the uncoupled sets of orthonormal pairs
\begin{eqnarray} &
\big\{h_1^+h_2^-, h_1^-h_2^+, h_1^0 h_2^0, h_1^0 I_2^0, I_1^0 h_2^0, I_1^0 I_2^0\big\} , ~~~ ~~
\big\{ h_1^+ h_2^0, h_1^+ I_2^0, h_2^+ h_1^0, h_2^+ I_1^0 \big\} ,
\nonumber \\ &
\Big\{ h_1^+ h_1^-, h_2^+ h_2^-, \tfrac{1}{\sqrt2}h_1^0 h_1^0, \tfrac{1}{\sqrt2} h_2^0 h_2^0,
\tfrac{1}{\sqrt2}I_1^0 I_1^0, \tfrac{1}{\sqrt2}I_2^0 I_2^0, \tfrac{1}{\sqrt2} D D\Big\} , ~~~~~
\big\{ h_1^0D,\, h_2^0D \big\} ~~~~~
\end{eqnarray}
to construct the matrix containing the tree-level amplitudes for \,$s_1^{}s_2^{}\to s_3^{}s_4^{}$,\,
which at high energies are dominated by the contributions of the four-particle contact diagrams.
We can write the distinct eigenvalues of this matrix as
\begin{eqnarray}
b_\pm^{} &=& \tfrac{1}{2}(\lambda_1+\lambda_2) \pm
\sqrt{\tfrac{1}{4}(\lambda_1-\lambda_2)\raisebox{0.3pt}{$^2$}+\lambda_4^2} \,, ~~~ ~~~~
c_\pm^{} \,=\, \tfrac{1}{2}(\lambda_1+\lambda_2) \pm
\sqrt{\tfrac{1}{4}(\lambda_1-\lambda_2)\raisebox{0.3pt}{$^2$}+\lambda_5^2} \,,
\nonumber \\
{\tt E}_\pm^{} &=& \lambda_3^{} + 2 \lambda_4^{} \pm 3 \lambda_5^{} \,, ~~~ ~~
{\tt F}_\pm^{} \,=\, \lambda_3^{} \pm \lambda_4^{} \,, ~~~ ~~
{\tt G}_\pm \,=\, \lambda_3^{} \pm \lambda_5^{} \,, ~~~ ~~
d_r^{} \,=\, 2\lambda_{rD}^{} \,, ~~~ r \,=\, 1,2 \,, ~~~ ~~
\end{eqnarray}
and the 3 solutions $a_{1,2,3}^{}$ of the cubic polynomial equation
\begin{eqnarray}
0 &=& a^3 - 3(\lambda_1+\lambda_2+\lambda_D) a^2 +
\big[ 9\lambda_1\lambda_2 - (2\lambda_3+\lambda_4)^2 + 9(\lambda_1+\lambda_2)\lambda_D
- 4 \lambda_{1D}^2 - 4 \lambda_{2D}^2 \big] a
\nonumber \\ && \!\! +\;
4 \big[ 3 \lambda_1^{} \lambda_{2D}^2 + 3 \lambda_2^{} \lambda_{1D}^2
- 2 (2 \lambda_3 + \lambda_4) \lambda_{1D}\lambda_{2D} \big] +
3 \big[ (2 \lambda_3 + \lambda_4)^2 - 9 \lambda_1 \lambda_2 \big] \lambda_D \,.
\end{eqnarray}
These results are consistent with those of Ref.\,\cite{Drozd:2014yla}.
The unitarity requirement for the \,$s_1^{}s_2^{}\to s_3^{}s_4^{}$\, amplitudes
then translates into the constraints
\begin{eqnarray} \label{unitarity}
|a_{1,2,3}|, |b_\pm|, |c_\pm|, |d_{1,2}|, |{\tt E}_\pm|, |{\tt F}_\pm|, |{\tt G}_\pm|
\,\le\, 8\pi \,.
\end{eqnarray}

The analogous conditions in the SM+D can be deduced from the foregoing by taking
the one-Higgs-doublet limit.
Thus, in the SM+D perturbativity demands \,$|\lambda_{\bar{\texttt H}}|\le8\pi$\, for the Higgs
self-coupling, \,$|\lambda_D|\le4\pi$,\, and \,$|\lambda|\le4\pi$,\,
whereas from Eq.\,(\ref{stability}) we have
\begin{eqnarray} \label{smd-stability}
\lambda_D^{} \,>\, 0 \,, ~~~ ~~~~ \lambda_{\bar{\texttt H}}^{} \,>\, 0 \,, ~~~ ~~~~
\lambda \,>\, -\sqrt{\tfrac{1}{2}\lambda_D^{}\lambda_{\bar{\texttt H}}^{}}
\end{eqnarray}
and from Eq.\,(\ref{unitarity})
\begin{eqnarray} \label{smd-unitarity}
\Big|\tfrac{3}{2}(\lambda_D+\lambda_{\bar{\texttt H}}) \pm
\sqrt{\tfrac{9}{4}(\lambda_D-\lambda_{\bar{\texttt H}})^2+4\lambda^2}\Big|
\,\le\, 8\pi \,, ~~~~ ~~~
|\lambda_{\bar{\texttt H}}| \,\le\, 8\pi \,, ~~~ ~~~~ |\lambda| \,\le\, 4\pi \,.
\end{eqnarray}
The first inequalities in the last line imply the stronger caps
\,$\lambda_{\bar{\texttt H}}\le8\pi/3$\, and \,$\lambda_D\le8\pi/3$.\,
The values of $|\lambda|$ shown in Fig.\,\ref{sm+d-plots}(a) are consistent with its
limit in Eq.\,(\ref{smd-unitarity}).

To implement the conditions in Eqs.\,\,(\ref{stability}) and (\ref{unitarity}),
we employ the relations
\begin{eqnarray} \label{lambda1}
\lambda_1^{} &=& \frac{s_\alpha^2 m_h^2+c_\alpha^2 m_H^2}{c_\beta^{2\,}v^2}
- \frac{s_\beta^{} m_{12}^2}{c_\beta^{3\,}v^2} \,, \hspace{9.93em}
\lambda_2^{} \,=\, \frac{c_\alpha^2 m_h^2+s_\alpha^2 m_H^2}{s_\beta^{2\,}v^2}
- \frac{c_{\beta\,}^{}m_{12}^2}{s_\beta^{3\,}v^2} \,, ~~~~~
\nonumber \\
\lambda_3^{} &=& \frac{s_{2\alpha}^{}}{s_{2\beta}^{}}~\frac{m_H^2-m_h^2}{v^2}
+ \frac{2m_{H^\pm}^2}{v^2} - \frac{2 m_{12}^2}{s_{2\beta}^{}v^2} \,, \hspace{6.27em}
\lambda_4^{} \,=\, \frac{m_A^2-2m_{H^\pm}^2}{v^2} +
\frac{2 m_{12}^2}{s_{2\beta}^{}v^2} \,, ~~~ ~
\nonumber \\
\lambda_5^{} &=& \frac{2 m_{12}^2}{s_{2\beta}^{}v^2} - \frac{m_A^2}{v^2} \,, ~~~~ ~~~
\lambda_{1D}^{} \,=\,
\frac{c_\alpha^{}\lambda_H^{}-s_\alpha^{}\lambda_h^{}}{c_\beta^{}} \,,  ~~~~ ~~~
\lambda_{2D}^{} \,=\, \frac{c_\alpha^{}\lambda_h^{}+s_\alpha^{}\lambda_H^{}}{s_\beta^{}} \,,
\end{eqnarray}
derived from ${\cal V}_{H,D}$.
Once $\alpha$ and $\beta$ have been specified, $m_{h,H,A,H^\pm,12}$ and $\lambda_{h,H}$ can
then serve as the free parameters instead of $\lambda_{1,2,3,4,5,1D,2D}$, as in
Eqs.\,\,(\ref{lambdahh}) and (\ref{lambdas}).
The expressions for $\lambda_{1,2,3,4,5}$ in Eq.\,(\ref{lambda1}) agree with those in
the literature~\cite{Gunion:2002zf}.


\begin{thebibliography}{0}

\bibitem{pdg}
  K.A.~Olive {\it et al.} [Particle Data Group Collaboration],
  Chin.\ Phys.\ C {\bf 38}, 090001 (2014)
and 2015 update.

\bibitem{lux} 
  D.S.~Akerib {\it et al.},
  arXiv:1608.07648 [astro-ph.CO].

\bibitem{pandax} 
  A.~Tan {\it et al.} [PandaX-II Collaboration],
Phys.\ Rev.\ Lett.\  {\bf 117}, no. 12, 121303 (2016)
  [arXiv:1607.07400 [hep-ex]].

\bibitem{dama} 
  R.~Bernabei {\it et al.},
  Eur.\ Phys.\ J.\ C {\bf 73}, 2648 (2013)  [arXiv:1308.5109 [astro-ph.GA]].

\bibitem{cdmssi} 
  R.~Agnese {\it et al.} [CDMS Collaboration],
  Phys.\ Rev.\ Lett.\  {\bf 111}, no. 25, 251301 (2013)  [arXiv:1304.4279 [hep-ex]].

\bibitem{cogent} 
  C.E.~Aalseth {\it et al.} [CoGeNT Collaboration],
  Phys.\ Rev.\ D {\bf 88}, 012002 (2013)  [arXiv:1208.5737 [astro-ph.CO]].

\bibitem{cresst} 
G.~Angloher {\it et al.},
  Eur.\ Phys.\ J.\ C {\bf 72}, 1971 (2012)  [arXiv:1109.0702 [astro-ph.CO]].

\bibitem{nosignal}
  J.H.~Davis, C.~McCabe, and C.~Boehm,
  JCAP {\bf 1408}, 014 (2014)  [arXiv:1405.0495]; 
  G.~Angloher {\it et al.} [CRESST-II Collaboration],
  Eur.\ Phys.\ J.\ C {\bf 74}, no. 12, 3184 (2014)  [arXiv:1407.3146]; 
  J.H.~Davis,
  Int.\ J.\ Mod.\ Phys.\ A {\bf 30}, no. 15, 1530038 (2015)  [arXiv:1506.03924]. 

\bibitem{Agnese:2014aze}
  R.~Agnese {\it et al.} [Super\-CDMS Collaboration],
  Phys.\ Rev.\ Lett.\  {\bf 112}, no. 24, 241302 (2014)  [arXiv:1402.7137 [hep-ex]].

\bibitem{Agnese:2015nto}
  R. Agnese {\it et al.} [Super\-CDMS Collaboration],
  Phys.\ Rev.\ Lett.\  {\bf 116}, no. 7, 071301 (2016)  [arXiv:1509.02448 [astro-ph.CO]].

\bibitem{Angloher:2015ewa}
  G. Angloher {\it et al.} [CRESST Collaboration],
  Eur. Phys. J. C {\bf 76}, no. 1, 25 (2016)  [arXiv:1509.01515 [astro-ph.CO]].

\bibitem{Cushman:2013zza}
  P.~Cushman {\it et al.},
  arXiv:1310.8327 [hep-ex].

\bibitem{xenon1t} 
  E. Aprile {\it et al.} [XENON Collaboration],
  JCAP {\bf 1604}, no. 04, 027 (2016)  [arXiv:1512.07501 [physics.ins-det]].

\bibitem{dsg2} 
  C.E.~Aalseth {\it et al.},
  Adv.\ High Energy Phys.\  {\bf 2015}, 541362 (2015).
  doi:10.1155/2015/541362

\bibitem{lz} 
  D.S.~Akerib {\it et al.} [LZ Collaboration],
  arXiv:1509.02910 [physics.ins-det].

\bibitem{nubg} 
  J. Billard, L. Strigari, and E. Figueroa-Feliciano,
  Phys.\ Rev.\ D {\bf 89}, no. 2, 023524 (2014)  [arXiv:1307.5458 [hep-ph]].

\bibitem{Kurylov:2003ra}
  A.~Kurylov and M.~Kamionkowski,
  Phys.\ Rev.\  D {\bf 69}, 063503 (2004)  [arXiv:hep-ph/0307185];
  F.~Giuliani,
  Phys.\ Rev.\ Lett.\  {\bf 95}, 101301 (2005)  [arXiv:hep-ph/0504157].

\bibitem{Feng:2011vu}
  J.L.~Feng, J.~Kumar, D.~Marfatia, and D.~Sanford,
  Phys.\ Lett.\  B {\bf 703}, 124 (2011)  [arXiv:1102.4331]. 

\bibitem{Feng:2013fyw}
  J.L.~Feng, J.~Kumar, and D.~Sanford,
  Phys.\ Rev.\ D {\bf 88}, no. 1, 015021 (2013)  [arXiv:1306.2315]. 

\bibitem{Savage:2008er}
  C. Savage, G. Gelmini, P. Gondolo, and K. Freese,
  JCAP {\bf 0904}, 010 (2009)  [arXiv:0808.3607]. 

\bibitem{inelastic} 
  N.~Chen, Q.~Wang, W.~Zhao, S.T.~Lin, Q.~Yue, and J.~Li,
  Phys.\ Lett.\ B {\bf 743}, 205 (2015)  [arXiv:1404.6043]; 
  C.Q.~Geng, D.~Huang, C.H.~Lee, and Q.~Wang,
  JCAP {\bf 1608}, no. 08, 009 (2016)  [arXiv:1605.05098 [hep-ph]].

\bibitem{Scopel:2015eoh}
  S.~Scopel and K.H.~Yoon,
  JCAP {\bf 1602}, no. 02, 050 (2016)  [arXiv:1512.00593 [astro-ph.CO]].

\bibitem{Silveira:1985rk}
  V.~Silveira and A.~Zee,
  Phys.\ Lett.\  B {\bf 161}, 136 (1985).

\bibitem{sm+reald}
  C.P.~Burgess, M.~Pospelov, and T.~ter Veldhuis,
  Nucl.\ Phys.\  B {\bf 619}, 709 (2001) [arXiv:hep-ph/0011335];
  M.C.~Bento {\it et al}., 
  Phys.\ Rev.\ D {\bf 62}, 041302 (2000)  [astro-ph/0003350];
  M.C.~Bento, O.~Bertolami, and R.~Rosenfeld,
  Phys.\ Lett.\ B {\bf 518}, 276 (2001)  [hep-ph/0103340];
  D.E.~Holz and A.~Zee,
  Phys.\ Lett.\ B {\bf 517}, 239 (2001)  [hep-ph/0105284];
  H.~Davoudiasl {\it et al}., 
  Phys.\ Lett.\  B {\bf 609}, 117 (2005)  [arXiv:hep-ph/0405097];
V.~Barger {\it et al}., 
  Phys.\ Rev.\  D {\bf 77}, 035005 (2008)  [arXiv:0706.4311]; 
  S.~Andreas, T.~Hambye, and M.H.G.~Tytgat,
  JCAP {\bf 0810}, 034 (2008)    [arXiv:0808.0255]; 
  C.E.~Yaguna,
  JCAP {\bf 0903}, 003 (2009)  [arXiv:0810.4267]; 
  M.~Gonderinger {\it et al}., 
  JHEP {\bf 1001}, 053 (2010)  [arXiv:0910.3167]; 
X.G.~He {\it et al}., 
  Phys.\ Lett.\  B {\bf 688}, 332 (2010)  [arXiv:0912.4722]; 
  M.~Asano and R.~Kitano,
  Phys.\ Rev.\ D {\bf 81}, 054506 (2010)  [arXiv:1001.0486]; 
  S.~Andreas {\it et al}., 
  Phys.\ Rev.\  D {\bf 82}, 043522 (2010)  [arXiv:1003.2595]; 
  A.~Badin and A.A.~Petrov,
  Phys.\ Rev.\ D {\bf 82}, 034005 (2010)  [arXiv:1005.1277]; 
  W.L.~Guo and Y.L.~Wu,
  JHEP {\bf 1010}, 083 (2010)  [arXiv:1006.2518]; 
  Nucl.\ Phys.\ B {\bf 867}, 149 (2013)  [arXiv:1103.5606]; 
  S.~Profumo, L.~Ubaldi, and C.~Wainwright,
  Phys.\ Rev.\  D {\bf 82}, 123514 (2010)  [arXiv:1009.5377]; 
  A.~Abada, D.~Ghaffor, and S.~Nasri,
  Phys.\ Rev.\ D {\bf 83}, 095021 (2011)  [arXiv:1101.0365]; 
  G.~Cynolter, E.~Lendvai, and G.~Pocsik,
  Acta Phys.\ Polon.\ B {\bf 36}, 827 (2005)  [hep-ph/0410102];
  A.~Biswas and D.~Majumdar,
  Pramana {\bf 80}, 539 (2013)  [arXiv:1102.3024]; 
  K.~Ghosh {\it et al}., 
  Phys.\ Rev.\  D {\bf 84}, 015017 (2011)  [arXiv:1105.5837]; 
  Y.~Mambrini,
  Phys.\ Rev.\  D {\bf 84}, 115017 (2011)  [arXiv:1108.0671]; 
  M.~Raidal and A.~Strumia,
  Phys.\ Rev.\  D {\bf 84}, 077701 (2011)    [arXiv:1108.4903]; 
  I.~Low {\it et al}., 
  Phys.\ Rev.\  D {\bf 85}, 015009 (2012)  [arXiv:1110.4405]; 
  O.~Lebedev, H.M.~Lee, and Y.~Mambrini,
  Phys.\ Lett.\  B {\bf 707}, 570 (2012)  [arXiv:1111.4482]; 
  A.~Drozd, B.~Grzadkowski, and J.~Wudka,
JHEP {\bf 1204}, 006 (2012) [{\it Erratum ibid.} {\bf 1411}, 130 (2014)] [arXiv:1112.2582];
  A.~Djouadi {\it et al}., 
  Phys.\ Lett.\ B {\bf 709}, 65 (2012)  [arXiv:1112.3299]; 
  Y.~Mambrini {\it et al}., 
  JCAP {\bf 1211}, 038 (2012)  [arXiv:1206.2352]; 
  K.~Cheung {\it et al}., 
  JCAP {\bf 1210}, 042 (2012)  [arXiv:1207.4930]; 
  L.B.~Jia and X.~Q.~Li,
  Phys.\ Rev.\ D {\bf 89}, no. 3, 035006 (2014)  [arXiv:1309.6029]; 
  F.S.~Queiroz, K.~Sinha, and A.~Strumia,
  Phys.\ Rev.\ D {\bf 91}, no. 3, 035006 (2015)  [arXiv:1409.6301]; 
  L.~Feng, S.~Profumo, and L.~Ubaldi,
  JHEP {\bf 1503}, 045 (2015)  [arXiv:1412.1105]; 
  M.~Duerr, 
  Phys.\ Lett.\ B {\bf 751}, 119 (2015)  [arXiv:1508.04418]; 
  JHEP {\bf 1606}, 152 (2016)  [arXiv:1509.04282]; 
  JHEP {\bf 1606}, 008 (2016)  [arXiv:1510.07562]; 
  H.~Han and S.~Zheng,
  JHEP {\bf 1512}, 044 (2015)  [arXiv:1509.01765]; 
  H.~Han {\it et al}., 
  Phys.\ Lett.\ B {\bf 756}, 109 (2016)  [arXiv:1601.06232]; 
  F.S.~Sage and R.~Dick,
  arXiv:1604.04589 [astro-ph.HE].

\bibitem{Cline:2013gha}
  J.M.~Cline {\it et al}., 
Phys.\ Rev.\ D {\bf 88}, 055025 (2013); {\bf 92}, no. 3, 039906(E) (2015)
  [arXiv:1306.4710]. 

\bibitem{sm+complexd}
  J.~McDonald,
  Phys.\ Rev.\  D {\bf 50}, 3637 (1994)  [arXiv:hep-ph/0702143];
V.~Barger, M.~McCaskey, and G.~Shaughnessy,
  Phys.\ Rev.\  D {\bf 82}, 035019 (2010)  [arXiv:1005.3328]; 
J.K.~Mizukoshi {\it et al}., 
  Phys.\ Rev.\ D {\bf 83}, 065024 (2011)  [arXiv:1010.4097]; 
  M.~Gonderinger, H.~Lim, and M.J.~Ramsey-Musolf,
  Phys.\ Rev.\ D {\bf 86}, 043511 (2012)  [arXiv:1202.1316]; 
  C.W.~Chiang, T.~Nomura, and J.~Tandean,
  Phys.\ Rev.\ D {\bf 87}, no. 7, 073004 (2013)  [arXiv:1205.6416]; 
  R.~Coimbra, M.O.P.~Sampaio, and R.~Santos,
  Eur.\ Phys.\ J.\ C {\bf 73}, 2428 (2013)  [arXiv:1301.2599]; 
  S.~Baek, P.~Ko, and W.I.~Park,
  Phys.\ Rev.\ D {\bf 90}, no. 5, 055014 (2014)  [arXiv:1405.3530]; 
R.~Costa {\it et al}., 
  Phys.\ Rev.\ D {\bf 92}, 025024 (2015)  [arXiv:1411.4048]; 
R.~Costa {\it et al}., 
  JHEP {\bf 1606}, 034 (2016)  [arXiv:1512.05355]. 

\bibitem{Bird:2006jd}
C.~Bird, R.~Kowalewski, and M.~Pospelov,
  Mod.\ Phys.\ Lett.\  A {\bf 21}, 457 (2006)  [arXiv:hep-ph/0601090];
X.G.~He {\it et al}., 
  Mod.\ Phys.\ Lett.\  A {\bf 22}, 2121 (2007) [arXiv:hep-ph/0701156].

\bibitem{He:2011gc}
  X.G.~He, B.~Ren, and J.~Tandean,
  Phys.\ Rev.\ D {\bf 85}, 093019 (2012)  [arXiv:1112.6364 [hep-ph]];
  X.G.~He and J.~Tandean,
  Phys.\ Rev.\ D {\bf 88}, 013020 (2013)  [arXiv:1304.6058 [hep-ph]].

\bibitem{He:2008qm}
  X.G.~He {\it et al}., 
  Phys.\ Rev.\ D {\bf 79}, 023521 (2009)  [arXiv:0811.0658 [hep-ph]].

\bibitem{2hdm+d}
  B.~Grzadkowski and P.~Osland,
  Phys.\ Rev.\  D {\bf 82}, 125026 (2010)  [arXiv:0910.4068]; 
  M.~Aoki, S.~Kanemura, and O.~Seto,
  Phys.\ Lett.\  B {\bf 685}, 313 (2010)  [arXiv:0912.5536]; 
  T.~Li and Q.~Shafi,
  Phys.\ Rev.\ D {\bf 83}, 095017 (2011)  [arXiv:1101.3576]; 
  Y.~Cai, X.G.~He, and B.~Ren,
  Phys.\ Rev.\  D {\bf 83}, 083524 (2011)  [arXiv:1102.1522]; 
  Y.~Bai {\it et al}., 
  Phys.\ Rev.\ D {\bf 88}, 015008 (2013)  [arXiv:1212.5604]; 
  Y.~Cai and T.~Li,
  Phys.\ Rev.\ D {\bf 88}, no. 11, 115004 (2013)  [arXiv:1308.5346]; 
  A.~Greljo {\it et al}., 
  JHEP {\bf 1311}, 190 (2013)  [arXiv:1309.3561]; 
  L.~Wang and X.F.~Han,
  Phys.\ Lett.\ B {\bf 739}, 416 (2014)  [arXiv:1406.3598]; 
  N.~Okada and O.~Seto,
  Phys.\ Rev.\ D {\bf 90}, no. 8, 083523 (2014)  [arXiv:1408.2583]; 
R.~Campbell {\it et al}., 
  Phys.\ Rev.\ D {\bf 92}, no. 5, 055031 (2015)  [arXiv:1505.01793]; 
  A.~Drozd {\it et al}., 
  JCAP {\bf 1610}, no. 10, 040 (2016)  [arXiv:1510.07053]. 

\bibitem{Drozd:2014yla}
  A.~Drozd {\it et al}., 
  JHEP {\bf 1411}, 105 (2014)  [arXiv:1408.2106]; 

\bibitem{lhc:mh}
  G.~Aad {\it et al.}  [ATLAS and CMS Collaborations],
  Phys.\ Rev.\ Lett.\  {\bf 114}, 191803 (2015)  [arXiv: 1503.07589 [hep-ex]].

\bibitem{lhctwiki}
  S.~Heinemeyer {\it et al.}  [LHC Higgs Cross Section Working Group Collaboration],
  arXiv:1307.1347 [hep-ph].
Online updates available at \\
https://twiki.cern.ch/twiki/bin/view/LHCPhysics/CERNYellowReportPageBR2014.

\bibitem{atlas+cms}
The ATLAS and CMS Collaborations,
  JHEP {\bf 1608}, 045 (2016)  [arXiv:1606.02266 [hep-ex]].

\bibitem{He:2010nt}
  X.G. He, S.Y. Ho, J. Tandean, and H.C. Tsai,
  Phys.\ Rev.\  D {\bf 82}, 035016 (2010) [arXiv:1004.3464 [hep-ph]];
  X.G.~He and J.~Tandean,
  Phys.\ Rev.\  D {\bf 84}, 075018 (2011)  [arXiv:1109.1277 [hep-ph]].

\bibitem{Cheng:2012qr}
  H.Y.~Cheng and C.W.~Chiang,
  JHEP {\bf 1207}, 009 (2012)  [arXiv:1202.1292 [hep-ph]].

\bibitem{thdm}
J.F.~Gunion, H.E.~Haber, G.L.~Kane, and S.~Dawson,
{\it The Higgs Hunter's Guide} (Westview Press, Colorado, 2000).

\bibitem{Branco:2011iw}
For a recent review, see
G.C.~Branco, P.M.~Ferreira, L.~Lavoura, M.N.~Rebelo, M.~Sher, and J.P.~Silva,
  Phys.\ Rept.\  {\bf 516}, 1 (2012)  [arXiv:1106.0034 [hep-ph]].

\bibitem{Shifman:1978zn}
  M.A.~Shifman, A.I.~Vainshtein, and V.I.~Zakharov,
  Phys.\ Lett.\  B {\bf 78}, 443 (1978);
  T.P.~Cheng,
  Phys.\ Rev.\  D {\bf 38}, 2869 (1988);
  H.Y.~Cheng,
  Phys.\ Lett.\  B {\bf 219}, 347 (1989).

\bibitem{isotopes}
J. Meija {\it et al}.,
Pure Appl. Chem. {\bf 88}, no. 3, 293 (2016).

\bibitem{Chen:2013vi}
C.S. Chen, C.Q. Geng, D. Huang, and L.H. Tsai,
Phys.\ Rev.\ D {\bf 87}, 075019 (2013) [arXiv:1301.4694]. 

\bibitem {Peskin:1991sw}
M.E. Peskin and T. Takeuchi, 
Phys.\ Rev.\ D {\bf 46}, 381 (1992).

\bibitem{Grimus:2008nb}
W. Grimus, L. Lavoura, O.M. Ogreid, and P. Osland,
Nucl.\ Phys.\ B {\bf 801}, 81 (2008) [arXiv:0802.4353]. 

\bibitem{Gondolo:1990dk}
  P.~Gondolo and G.~Gelmini,
  Nucl.\ Phys.\ B {\bf 360}, 145 (1991).

\bibitem{Kolb:1990vq}
E.W. Kolb and M. Turner, {\it The Early Universe} (Westview Press, Boulder, 1990).

\bibitem{Steigman:2012nb}
  G.~Steigman, B.~Dasgupta, and J.F.~Beacom,
  Phys.\ Rev.\ D {\bf 86}, 023506 (2012)  [arXiv:1204.3622]. 

\bibitem{planck} 
  P.A.R.~Ade {\it et al.} [Planck Collaboration],
  Astron.\ Astrophys.\  {\bf 594}, A13 (2016)  [arXiv:1502.01589 [astro-ph.CO]].

\bibitem{Kanemura:1999xf}
  S.~Kanemura, T.~Kasai, and Y.~Okada,
  Phys.\ Lett.\ B {\bf 471}, 182 (1999)  [hep-ph/9903289].

\bibitem{Kanemura:1993hm}
  S.~Kanemura, T.~Kubota, and E.~Takasugi,
  Phys.\ Lett.\ B {\bf 313}, 155 (1993)  [hep-ph/9303263].

\bibitem{Akeroyd:2000wc}
A.G.~Akeroyd, A.~Arhrib, and E.M.~Naimi,
Phys.\ Lett.\ B {\bf 490}, 119 (2000)  [hep-ph/0006035].

\bibitem{copositivity}
K. Hadeler, Linear Algebra Appl. {\bf 49}, 79 (1983);
G.~Chang and T.W.~Sederberg, Comput. Aided Geom. Des. {\bf 11} (1), 113 (1994);
L.~Ping and F.Y.~Yu, Linear Algebra Appl. {\bf 194}, 109 (1993);
  K.~Kannike,
  Eur.\ Phys.\ J.\ C {\bf 72}, 2093 (2012)  [arXiv:1205.3781 [hep-ph]].

\bibitem{Gunion:2002zf}
  J.F.~Gunion and H.E.~Haber,
  Phys.\ Rev.\ D {\bf 67}, 075019 (2003)  [hep-ph/0207010].


\end{thebibliography}
\end{document}